\documentclass[%
 reprint,
 amsmath,amssymb,
 aps,
 showkeys,
]{revtex4-2}

\usepackage{graphicx}
\usepackage{dcolumn}
\usepackage{bm}
\usepackage{amsmath}
\usepackage{xcolor}
\renewcommand{\eqref}[1]{Eq. \ref{#1}}
\newcommand{\figref}[1]{Fig. \ref{#1}}

\begin{document}

\title{Fluid drainage in erodible porous media}


\author{Joanna Schneider$^{1}$, Christopher A. Browne$^{1}$, Malcolm Slutzky$^{2}$, Cecilia A. Quirk$^{3}$, Daniel B. Amchin$^{1}$, Sujit S. Datta$^{1}$}

\email{ssdatta@princeton.edu}
\affiliation{%
 $^{1}$Department of Chemical and Biological Engineering, Princeton University, Princeton, NJ 08544\\
 $^{2}$Department of Physics, Princeton University, Princeton, NJ 08544\\
 $^{3}$Department of Operations Research and Financial Engineering, Princeton University, Princeton, NJ 08544}

\date{\today}

\begin{abstract}
Drainage, in which a nonwetting fluid displaces a wetting fluid from a porous medium, is well-studied for media with unchanging solid surfaces. However, many media can be eroded by drainage, with eroded material redeposited in pores downstream, altering further flow. Here, we use theory and simulation to examine how these coupled processes both alter the overall fluid displacement pathway and help reshape the solid medium. We find two new drainage behaviors with markedly different characteristics, and quantitatively delineate the conditions under which they arise. Our results thereby help expand current understanding of these rich physics, with implications for applications of drainage in industry and the environment.

\end{abstract}


\maketitle

Drainage is the process by which a nonwetting fluid displaces a wetting fluid from a porous
medium. It underlies a broad range of environmental and industrial processes, 
including groundwater contamination, oil migration and recovery, 
gas venting from sediments, CO$_2$ sequestration, soil drying, and fluid transport in porous membranes \cite{bethke1991long,kueper1991two,dawson1997influence,levy2003modelling,dandekar2006petroleum,benson2008carbon,cueto2008nonlocal,neufeld2009modelling,bear2010modeling,macminn2010co2,saadatpoor2010new,bandara2011pore,sahimi2011flow,berg2012stability,carmo2013comprehensive,lee2016porous,bazyar2018liquid}. Therefore, extensive research has sought to develop ways to predict 
the displacement pathway taken by the nonwetting fluid \cite{knackstedt2009invasion,blunt2017multiphase}, building on the seminal model of \textit{invasion percolation} proposed by Wilkinson and Willemsen four decades ago~\cite{wilkinson1983invasion}. 

In this model, the medium is assumed to be composed of a static solid matrix of uniform wettability (with a prescribed three-phase contact angle $\theta$) that houses an interconnected network of pores with randomly varying sizes. The nonwetting fluid is taken to be much more viscous than the wetting fluid, and its flow is considered to be very slow; in this limit, which characterizes many real-world processes, capillary forces at the immiscible fluid interface dictate the resulting displacement pathway. In particular, the nonwetting fluid cannot invade a pore of entrance radius $r$ until the capillary pressure difference across the interface reaches a threshold $\Delta p_c\equiv{2\gamma\cos\theta}/{r}$, where $\gamma$ is the interfacial tension between the two fluids. Hence, the fluid displacement proceeds one pore at a time---with the nonwetting fluid invading the largest pore accessible to it, and therefore the lowest capillary pressure threshold, successively. The fluid displacement pathway is then determined by random local variations in pore size, resulting in a characteristic ramified and disordered displacement pattern known as \textit{capillary fingering} (CF) \cite{mayer1965mercury,lenormand1983mechanisms,lenormand1985invasion,mason1986meniscus,lenormand1989capillary,martys1991critical,maaloy1992dynamics,toledo1994pore,xu1998invasion,xu2008dynamics,joekar2012analysis,krummel2013visualizing}.

While this foundational model has been validated in highly-controlled lab studies, it makes a strong assumption that often does not hold in practice: that the structure of the solid matrix is unchanging. In reality, capillary forces at the immiscible fluid interface can restructure the matrix. One way this can happen is by deforming or fracturing the overall medium \cite{holtzman2010crossover,holtzman2012capillary}. Another way is by eroding frangible \cite{derr2020flow} and plastocapillary \cite{style2015adsorption} material from the walls of the solid matrix and redepositing it within the pore space downstream. A prominent example is the layers of colloidal particles, inorganic precipitates, and organic matter that frequently coat the mineral grains making up soils, sediments, and subsurface aquifers/reservoirs \cite{means1982role,tipping1982effect,gibbs1983effect,corapcioglu1993colloid,ouyang1996colloid,hendraningrat2013laboratory,feia2015experimental,gerber2019self,bizmark2020multiscale,gerber2020propagation,li2020asphaltene}. Field observations indicate that fluid drainage caused by processes like wetting/drying cycles and contaminant/oil migration can erode and redeposit these materials, impacting subsequent transport over large scales \cite{mccarthy1989t,kan1990ground,johnson1996enhanced,franchi2003effects,pelley2008effect}. However, despite their common occurrence, the influence of solid erosion \& deposition on fluid drainage---and vice versa---has, to our knowledge, never been studied.
  
Here, we incorporate these new physics  into the classic framework of invasion percolation. Our numerical simulations reveal two new drainage behaviors whose fluid displacement and solid deposition patterns differ dramatically from standard CF: \textit{rapid clogging}, in which redeposited material rapidly clogs the pore space and arrests subsequent flow, and \textit{erosion-enhanced fingering}, in which constriction of some pores by deposition unexpectedly enables the nonwetting fluid to invade a greater fraction of the medium. Furthermore, we use calculations to delineate the conditions under which these different behaviors arise, governed by two dimensionless parameters that quantify how much of, and how easily, the solid matrix can be eroded. 

\begin{figure*}
    \centering
    \includegraphics[width=0.95\textwidth]{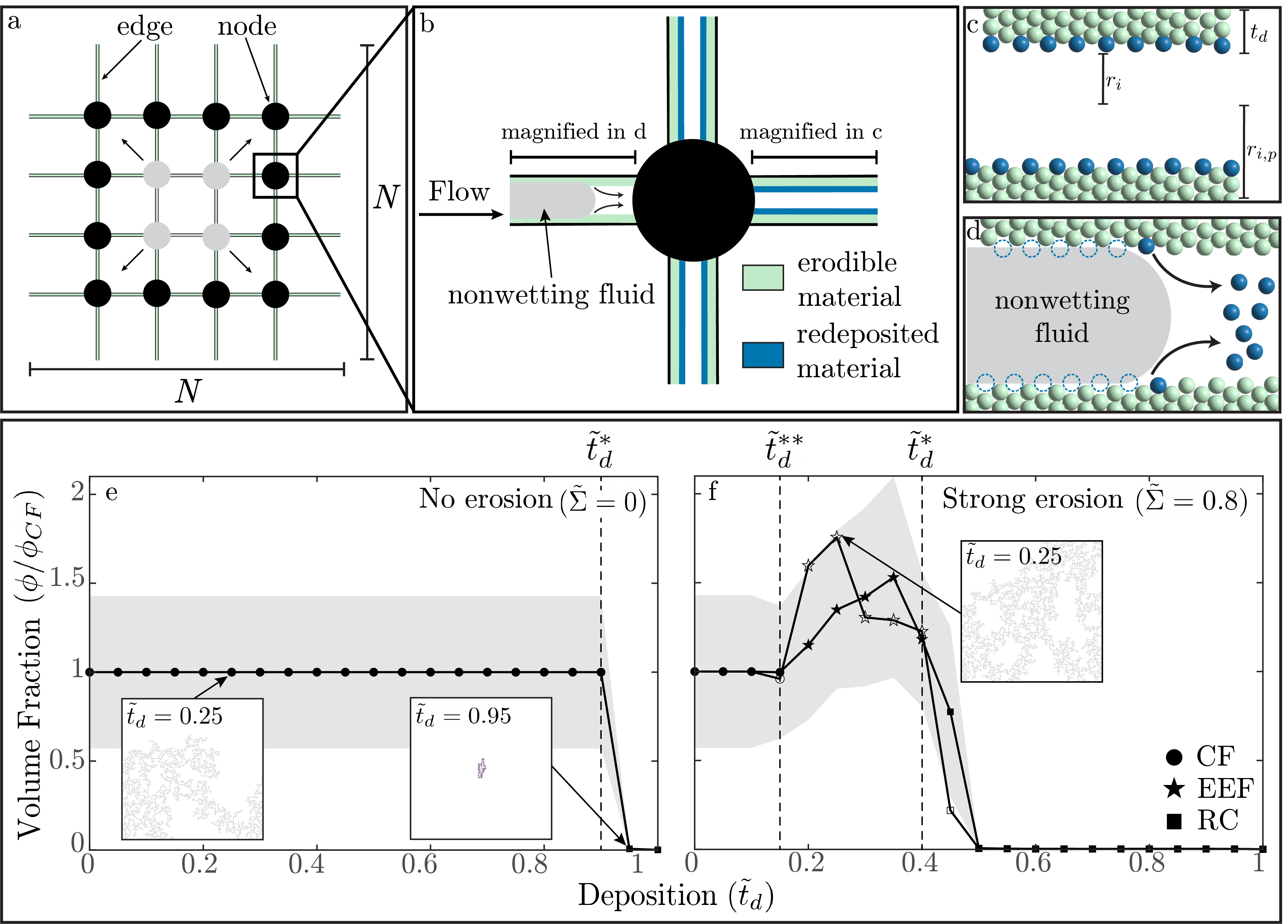}
    \caption{Network modeling of fluid drainage in an erodible porous medium reveals new drainage behaviors. (a) Schematic of the 2D network model, with $N \times N$ nodes representing the pore bodies, and the edges representing the interconnecting pore throats. (b) Magnified view of a single pore body. (c) Pore throats initially have a pristine radius $r_{i,p}$, with an initial layer (mint) of erodible material $t_d$ thick. (d) As the nonwetting fluid enters a pore during drainage, it erodes some of this material, redepositing it into connected throats (blue). Numerical simulations reveal new drainage behaviors arising from solid erosion \& deposition; we characterize these behaviors in (e--f) using the volume fraction of the pore space filled by the nonwetting fluid after drainage completes, $\phi$, normalized by the case of standard capillary fingering (CF), $\phi_{CF}$. Open symbols show the results for the single network used to generate the simulations shown in the insets. Closed symbols and gray shading show the average and standard deviation, respectively, of results obtained over 100 different, but statistically-identical, networks. (e) Without erosion ($\tilde{\Sigma} = 0$), drainage proceeds by CF (magnified view in left inset) until $\tilde{t}_d>\tilde{t}^{*}_d\approx0.9$, above which the medium starts with so much erodible material that pores near the inlet are clogged. The medium transitions to rapid clogging (RC), shown by the magnified view in the right inset; circular gray and purple markers indicate invaded and clogged pore throats, respectively, while gray $+$ symbols denote invaded pore bodies. (f) With strong erosion ($\tilde{\Sigma} = 0.8$), drainage proceeds by CF only until $\tilde{t}_d>\tilde{t}^{**}_d\approx0.15$, above which the nonwetting fluid unexpectedly explores \emph{more} of the pore space than in CF. Drainage proceeds by erosion-enhanced fingering (EEF), shown by the magnified view in the inset. With increasing $\tilde{t}_d>\tilde{t}^{*}_d\approx0.4$, clogging increasingly dominates, and drainage transitions back to RC.}
    \label{Figure 1}
\end{figure*}

\emph{\textbf{Model development.}} To begin to unravel the complex physics underlying this problem, we examine a simple, but illustrative, example. Following the typical approach of pore-network modeling \cite{wilkinson1986percolation,birovljev1991gravity,masson2016fast}, we consider fluid drainage in a 2D network of $N\times N$ nodes, which represent the pore ``bodies'', with locations defined by an adjacency matrix with network connectivity $c$ (\figref{Figure 1}a). The edges between nodes are indexed serially by $i$ and represent the constrictions (``throats'') between pores. Because $p_c$ is locally maximum at these constrictions, they control both pore invasion and solid erosion. We thus focus our attention on the edges of the network; for simplicity, we assume that they compose the entirety of the pore space volume, and approximate them as cylinders of uniform length $L$ and pristine radii $r_{i,p}$ drawn randomly from a given distribution $\rho(r_{i,p})$. 

To impart erodibility to this static matrix, we consider the inner wall of each pore throat to also be coated by a layer of erodible material, initially of constant thickness $t_d$ (\figref{Figure 1}b--c) distributed uniformly throughout the medium. The effective radius of throat $i$ is then given by $r_{i} = r_{i,p}-t_d$, with a corresponding capillary pressure threshold $\Delta p_{c}(r_i)={2\gamma\cos\theta}/{r_{i}}$; without loss of generality, we take $\theta=0$. Motivated by studies in single pores \cite{jager2017channelization,khodaparast2017bubble,yu2017armoring,yin2018dynamic, jeong2022particulate}, we account for drainage-induced erosion using a simple rule: if a pore is invaded by the nonwetting fluid, the moving immiscible fluid interface erodes material from the wall (\figref{Figure 1}d) when $\Delta p_{c}(r_i)$ exceeds a threshold stress $\sigma_{y}$ that quantifies the material's durability, analogous to a yield stress. For ease of notation, we indicate dimensionless quantities by overtildes $(^\sim)$, and nondimensionalize all length scales by $r_{p, \rm{max}}\equiv\max\{r_{i,p}\}$. The ratio $\tilde{\Sigma}\equiv\Delta p_{c}(r_{p,\rm{max}})/\sigma_{y}$ then compares the smallest capillary pressure that can possibly arise in the porous medium to the threshold erosion stress; that is, it describes the relative ease with which the immiscible fluid interface erodes material from the pore walls as it moves. We therefore call this dimensionless parameter the medium's \textit{erodibility}. 

Hence, as the nonwetting fluid invades a pore with throat radius $r_i$, the amount of material eroded from its walls depends on $\tilde{\Sigma}$. If $\tilde{\Sigma}<\tilde{r}_{i}$, erosion does not occur, and the dimensions of the pore remains unchanged after drainage; the radius after the entire drainage process has completed, $\tilde{r}'_{i}$, remains equal to $\tilde{r}_{i}$. Above the threshold $\tilde{\Sigma}\geq\tilde{r}_{i}$, erosion causes the radius to increase to a new value $r'_{i}=2\gamma/\sigma_{y}$ at which the corresponding capillary pressure becomes balanced by the threshold stress for erosion, or equivalently, $\tilde{r}'_{i}=\tilde{\Sigma}$. However, there is a limit to how much material can be eroded from a pore: if the erodibility is so large that this new value $2\gamma/\sigma_{y}$ exceeds the pristine radius $r_{i,p}$ (that is, if $\tilde{\Sigma}>\tilde{r}_{i,p}$), then the pore throat radius saturates at its largest possible value, $\tilde{r}'_{i}=\tilde{r}_{i,p}$.

Finally, we also incorporate the subsequent deposition of the eroded material in the non-drained throats $j$ directly connected to a drained, eroded throat $i$. In particular, because we assume cylindrical pore throats with $N\gg1$, we distribute the volume eroded from $i$ proportionately to $\sim r_{j}^{4}$, following mass conservation (detailed in \cite{SI}), reducing the values of $\tilde{r}'_{j}$ accordingly. However, if this process causes a pore throat $j$ to become fully clogged, the excess volume of eroded material is returned to the parent $i$, and the throat is removed from the network to prevent subsequent flow through it. 

\emph{\textbf{Model implementation.}} To characterize the influence of solid erosion \& deposition on fluid drainage, we perform numerical simulations of this model with $N=200$, $c=4$, and $\rho(\tilde{r}_{i,p})$ given by a uniform distribution spanning $\tilde{r}_{i,p}\in[0.83,1]$; we find similar results to those described below when exploring other values of $N$, $c$, and forms of $\rho(\tilde{r}_{i,p})$, including those obtained from real-world media \cite{SI}. For each simulation condition tested, parameterized by prescribed input values of $(\tilde{t}_d,\tilde{\Sigma})$, we run 100 unique iterations, each with $\tilde{r}_{i,p}$ randomly sampled from the same $\rho(\tilde{r}_{i,p})$. In each simulation, the pore bodies and throats all start saturated with the wetting fluid, and drainage is initiated by introducing the nonwetting fluid at the four central pore bodies \cite{SI}. During each time step, we then determine the connected component clusters of undrained pore bodies; the boundaries with these clusters delineate the invading nonwetting fluid interface or trapped wetting fluid regions. Following standard invasion percolation, we then identify the largest pore throat $i$, with the smallest capillary pressure threshold $\Delta p_{c}\sim1/\tilde{r}_{i}$, along the invading nonwetting fluid interface. We fill the corresponding pore throat and body, keeping trapped wetting fluid regions unchanged to model an incompressible fluid, and incorporating solid erosion \& deposition following the rules described above. We then iterate through time steps until the nonwetting fluid reaches the periphery of the network or is completely surrounded by clogged pores. 

\emph{\textbf{Solid erosion \& deposition engender fundamentally new drainage behaviors.}} As a baseline, we first establish the classic case of invasion percolation without any erosion $(\tilde{t}_d=0,\tilde{\Sigma}=0)$. As expected, drainage occurs through a series of successive bursts along a ramified, disordered pathway characteristic of typical CF (Movie S1). The resulting nonwetting fluid pathway fills a fraction $\phi=\phi_{CF}= 0.10\pm0.04$ of the total pore space volume and has a fractal dimension \cite{niemeyer1984fractal} $d_f = 1.86\pm0.04$, in good agreement with previous studies of CF~\cite{lenormand1985invasion,knackstedt2009invasion,blunt2017multiphase}. Furthermore, slightly increasing the amount of erodible material, but without any erosion $(0<\tilde{t}_d<0.9,\tilde{\Sigma}=0)$, still results in CF (\figref{Figure 1}e, left \& Movie S2)---as expected, since in this case, all pores are simply constricted uniformly. However, increasing further above a threshold value $\tilde{t}_d=\tilde{t}_d^*\approx0.9$ causes a precipitous drop in $\phi$ (\figref{Figure 1}e, right) as pores near the inlet clog, preventing fluid drainage from occurring (Movie S3). We therefore call this behavior \emph{rapid clogging} (RC).

Next, we explore the case of high erodibility ($\tilde{\Sigma} = 0.8$). When the amount of material that can be eroded is small ($\tilde{t}_d \leq0.1$), the influence of erosion \& deposition is minimal, and drainage again proceeds through typical CF (\figref{Figure 1}f, circles). We observe dramatically different behavior with increasing $\tilde{t}_d$. Above a threshold value $\tilde{t}_d=\tilde{t}_d^{**}\approx0.15$, the nonwetting fluid volume fraction is \emph{larger} than in CF ($\phi/\phi_{CF}>1$, \figref{Figure 1}f, stars)---that is, as \emph{more} erodible material is added to the pore space, the nonwetting fluid is somehow able to form new, ramified fingers and thereby drain \emph{more} of the pore space (\figref{Figure 1}f, inset \& Movie S4). We therefore call this behavior \emph{erosion-enhanced fingering} (EEF). This surprising behavior persists with increasing $\tilde{t}_d$ until it eventually becomes suppressed by pore clogging; in this case, we again observe a transition to RC, characterized by a precipitous drop in $\phi/\phi_{CF}$, above a threshold value $\tilde{t}_d=\tilde{t}_d^*\approx0.4$ (\figref{Figure 1}f, squares).

\begin{figure}
    \centering
    \includegraphics[width = 0.46\textwidth]{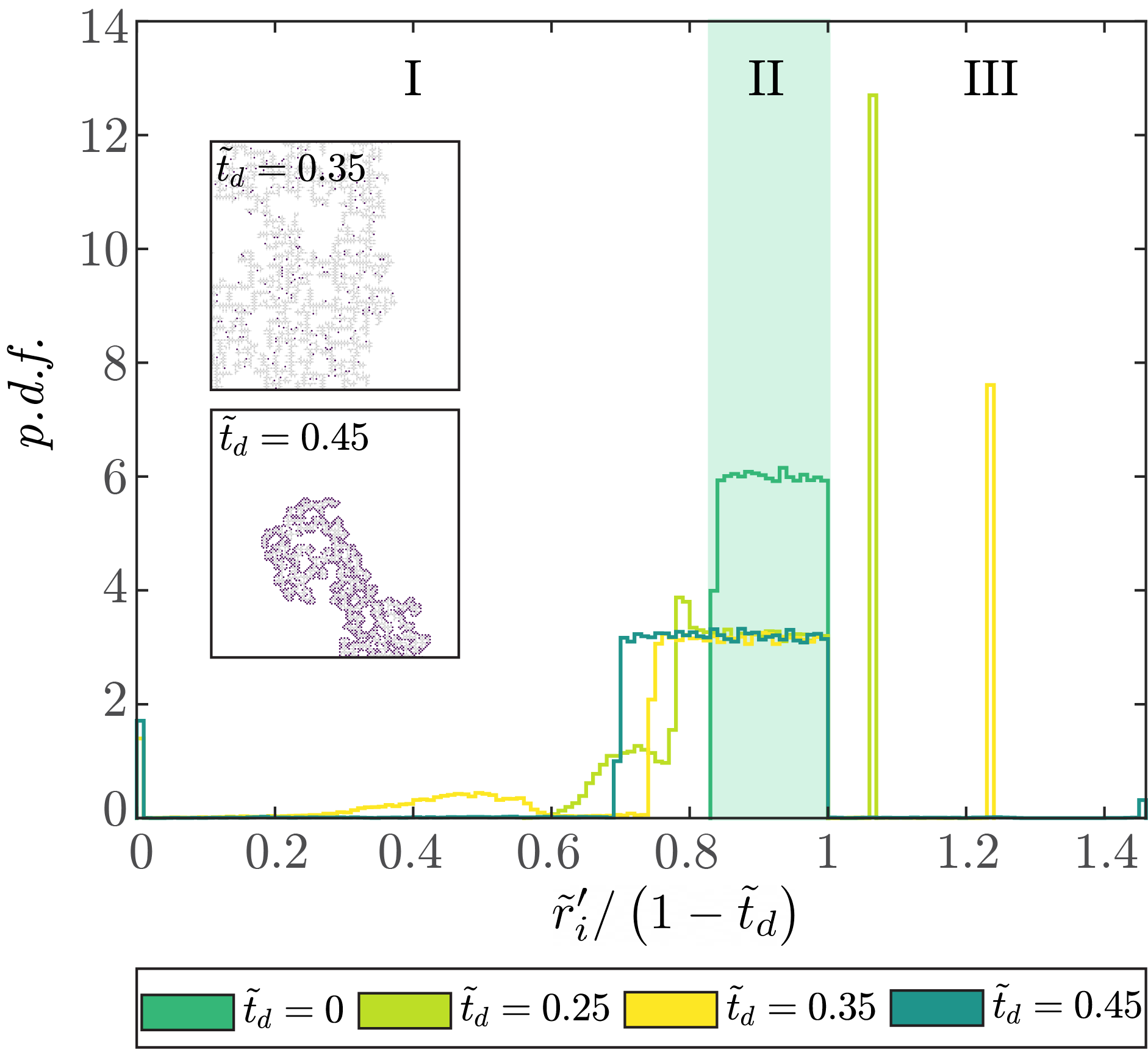}
    \caption{Examining the probability density function (\textit{p.d.f.}) of $\tilde{r}'_{i}/\left(1-\tilde{t}_d\right)$, the pore sizes after drainage relative to the largest starting pore size, elucidates the origins of new drainage behaviors. We consider the representative case of strong erosion ($\tilde{\Sigma}=0.8$) shown in \figref{Figure 1}f. The initial uniform distribution is shown in Region II for the pristine case without any erodible material ($\tilde{t}_d = 0$). Above the threshold to transition to EEF $\tilde{t}_d \approx 0.15$, two subfractions of smaller and larger pores (Regions I and III) split off---reflecting pores that have had material eroded from and redeposited in, respectively. At larger $\tilde{t}_d$ above the threshold $\tilde{t}_d \approx 0.4$, increasing clogging (peak in Region I) causes a transition to RC. Insets show magnified views of the resulting patterns of nonwetting fluid displacement (gray) and pore clogging (purple) for $\tilde{t}_d = 0.35$ and $0.45$.}
    \label{Figure 2}
\end{figure}

\begin{figure}
    \centering
    \includegraphics[width = 0.48\textwidth]{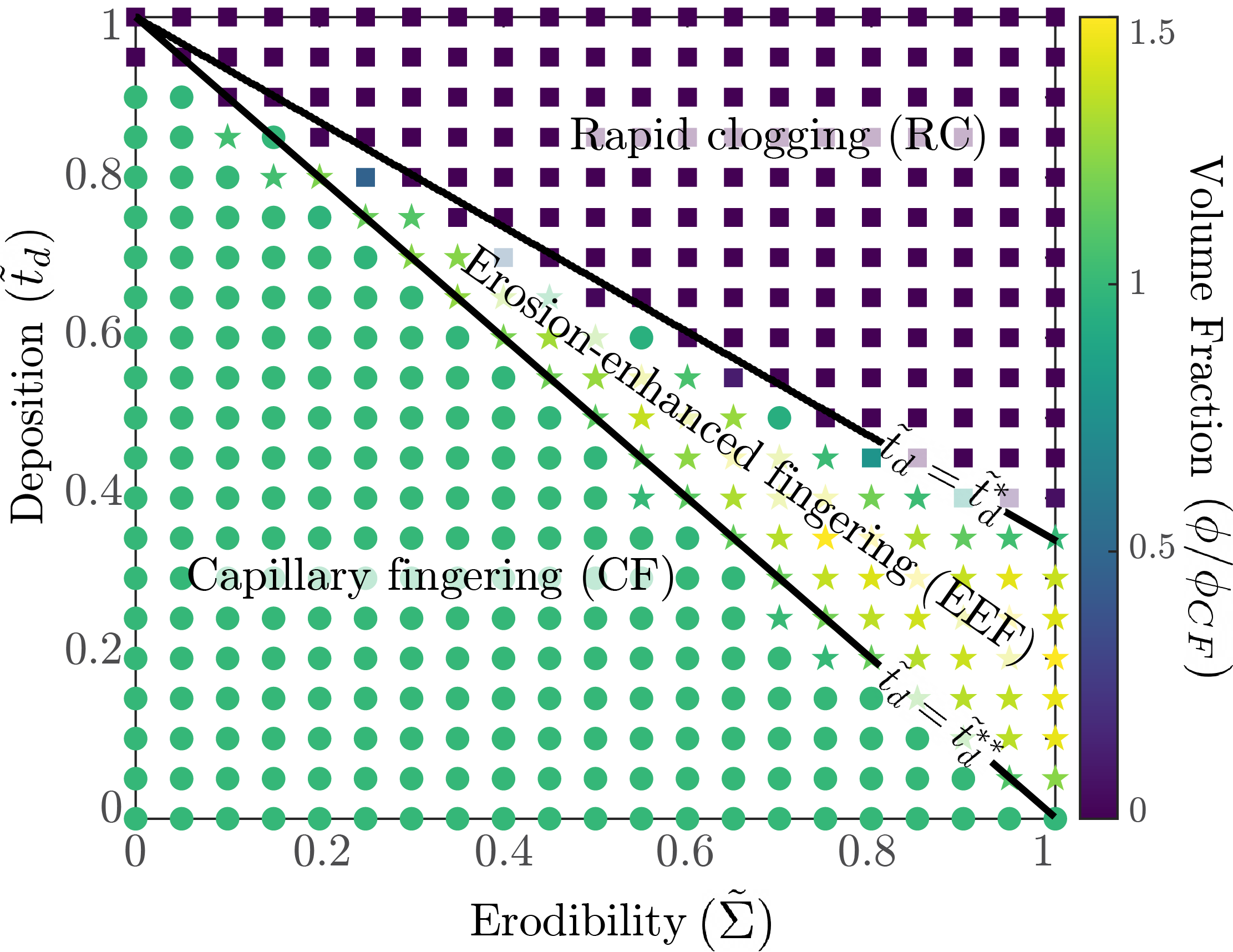}
    \caption{State diagram of different drainage behaviors in an erodible porous medium. Colors show the normalized nonwetting fluid volume fraction $\phi/\phi_{CF}$; each symbol shows the average of 100 different simulations testing different, but statistically-identical, networks. We observe the emergence of three distinct drainage behaviors: capillary fingering (circles, $0.9<\phi/\phi_{CF}<1$), rapid clogging (squares, $\phi/\phi_{CF}<0.9$), and erosion enhanced fingering (stars, $\phi/\phi_{CF}>1$); as shown in \cite{SI}, these behaviors can also be characterized by the distinct fractal dimensions of the resulting drainage patterns. 
}
    \label{Figure 3}
\end{figure}

\emph{\textbf{Origins of these new drainage behaviors.}} Why do these fascinating new drainage behaviors arise in erodible porous media? Inspecting changes in the distribution of pore sizes after drainage, which quantifies how the nonwetting fluid displacement has reshaped the pore space structure, provides a clue. In particular, we examine the distributions of $\tilde{r}'_{i}/\left(1-\tilde{t}_d\right)=\tilde{r}'_{i}/\tilde{r}_{i,\rm{max}}$, which describe the pore sizes after drainage relative to the largest starting pore size $\tilde{r}_{i,\rm{max}}\equiv\max\{\tilde{r}_{i}\}$. We focus on the highly erodible case of $\tilde{\Sigma}=0.8$ described in \figref{Figure 1}f as a representative example. When the medium only has a little erodible material ($\tilde{t}_d<0.15$), the initial uniform distribution of pore sizes remains unaltered (Region II in \figref{Figure 2}). However, as exemplified by $\tilde{t}_d=0.25$ in \figref{Figure 2}, just above the threshold $\tilde{t}_d^{**}\approx0.15$, two sub-fractions of smaller and larger pores (Regions I and III, respectively) split off from this distribution. These reflect the increasing fraction of pore throats that have had solid material eroded from and redeposited in, respectively; indeed, the eroded pores reach a uniform size set by the balance of capillarity and erosion, with $\tilde{r}'_{i}/\tilde{r}_{i,\rm{max}}\approx1.1=\tilde{\Sigma}/(1-\tilde{t}_d)$, as expected. Notably, the smaller pores still have sizes $\tilde{r}_i>0$, indicating that they have not yet reached the threshold for clogging. We observe similar behavior in \figref{Figure 2} for the cases of $\tilde{t}_d=0.35$ and $0.45$, for which the eroded pores now reach the expected sizes $\tilde{r}'_{i}/\tilde{r}_{i,\rm{max}}=\tilde{\Sigma}/(1-\tilde{t}_d)\approx1.2$ and $\approx1.5$, respectively.

Thus, we expect that EEF begins when capillary forces become just large enough to begin eroding the solid matrix---and the redeposition of this material constricts downstream pores slightly, just enough to force the nonwetting fluid to explore new pathways through the pore space that it otherwise would not have. We quantify this expectation for the onset of EEF by balancing the smallest capillary pressure that can possibly be encountered during drainage, $\Delta p_{c}\left(r_{p,\rm{max}}-t^{**}_{d}\right)$, with the threshold erosion stress $\sigma_{y}$. In nondimensional form, our prediction is:
\begin{equation}
\tilde{t}^{**}_{d}=1-\tilde{\Sigma}.
    \label{bottom boundary}
\end{equation}
This prediction yields  $\tilde{t}^{**}_{d}=0.2$, in good agreement with the value of $\tilde{t}^{**}_{d}\approx0.15$ found from the simulations for the case of $\tilde{\Sigma}=0.8$.

As $\tilde{t}_d$ increases above $\tilde{t}^{**}_{d}$, we expect that the increasing amount of erodible material increases the propensity of pores to become clogged---giving rise to the non-monotonic variation of $\phi$ shown in \figref{Figure 1}f. Consistent with this expectation, a larger fraction of pores in Region I becomes clogged (shown by the growing peak at $\tilde{r}'_{i}=0$, also indicated by the purple points in the insets to \figref{Figure 2} and Movies S5--S6). The height of the peak in Region III concomitantly decreases, indicating that fewer pores are ultimately eroded.

Thus, we expect that EEF transitions to RC when pore clogging is so prevalent that it ``chokes off" fluid drainage. We quantify this expectation for the onset of RC by balancing the volume of solid material that can be eroded from a pore $i$, $\propto\left(2\gamma/\sigma_y\right)^2-\left({r}_{i,p}-{t}^{*}_d\right)^2$, with the characteristic available volume in the adjacent connected pores $j$, $\propto \alpha\left({r}_{j,p}-{t}^{*}_d\right)^2$, where the constant $\alpha\approx4/3$ accounts for the network connectivity \cite{SI}. While both $r_{i,p}$ and $r_{j,p}$ are broadly distributed, we make the assumption that both are $\sim r_{p,\rm{max}}$. With this simplification, in nondimensional form, our prediction is:
\begin{equation}\tilde{t}^{*}_d=1-
\frac{\tilde{\Sigma}}{\sqrt{1+\alpha}}.
    \label{top boundary}
\end{equation}
This prediction yields  $\tilde{t}^{*}_{d}=0.4$, in excellent agreement with the value of $\tilde{t}^{*}_{d}\approx0.4$ found from the simulations, for the case of $\tilde{\Sigma}=0.8$.

\emph{\textbf{A unified state diagram for drainage in an erodible porous medium.}} As a final test of the predictions given by Eqs.~(\ref{bottom boundary}) \& (\ref{top boundary}), we perform a total of 44,100 numerical simulations over a broad range of $(\tilde{t}_d,\tilde{\Sigma})$. We characterize the drainage pattern that emerges for each condition tested using the volume fraction and fractal dimension \cite{niemeyer1984fractal} of the nonwetting fluid pathway, $\phi$ and $d_f$, respectively. Our results are summarized in \figref{Figure 3}. Consistent with the observations shown in Figs.~\ref{Figure 1}--\ref{Figure 2}, CF emerges for small $(\tilde{t}_d,\tilde{\Sigma})$ (circles), transitioning to EEF for $\tilde{t}_d\geq\tilde{t}^{**}_d$ (stars), and then transitioning to RC for $\tilde{t}_d\geq\tilde{t}^{*}_d$ (stars). Moreover, the boundaries between these distinct drainage behaviors agree well with the predictions given by Eqs.~(\ref{bottom boundary}) \& (\ref{top boundary}), shown by the lower and upper solid lines, respectively---despite the simplifying assumptions made therein. Thus, not only has our extended model of invasion percolation shown that the coupling between nonwetting fluid displacement and solid erosion \& deposition engender fascinating new drainage behaviors, but our analysis provides quantitative principles to help predict when they arise. Future work could build on the framework developed here by exploring a broader range of fluid viscosity ratios \cite{xu1998invasion} and flow rates (extending Lenormand's classic phase diagram \cite{lenormand1988numerical}), as well as different forms of pore space structure \cite{meakin1992invasion,onody1995experimental,al2012control,datta2013drainage,jackson2017stability,biswas2018drying,lu2019controlling,lu2020forced,lu2021forced}, and different rules for erosion, clogging, and potential clog erosion---ultimately leading to improved prediction and control of coupled fluid and solid transport in diverse environmental and industrial media. 

\begin{acknowledgments}
It is a pleasure to acknowledge I.C. Bourg, H.A. Stone, and S. Torquato for stimulating discussions, as well as N. Bizmark, E.Y. Chen, A. Hancock, and N. Subraveti for helpful feedback on the manuscript. This work was supported by funding from the New Jersey Water Resources Research Institute, the ReMatch+ program (to CAQ), a Mary and Randall Hack Graduate Award of the High Meadows Environmental Institute (to JS), a Maeder Graduate Fellowship from the Andlinger Center for Energy and the Environment (to JS), and the Princeton Center for Complex Materials (PCCM), a National Science Foundation (NSF) Materials Research Science and Engineering Center funded through NSF grant DMR-2011750.
\end{acknowledgments}

 \clearpage

\section{Supplementary Materials}

\begin{figure}[h]
    \centering
    \includegraphics[width = 0.25\textwidth]{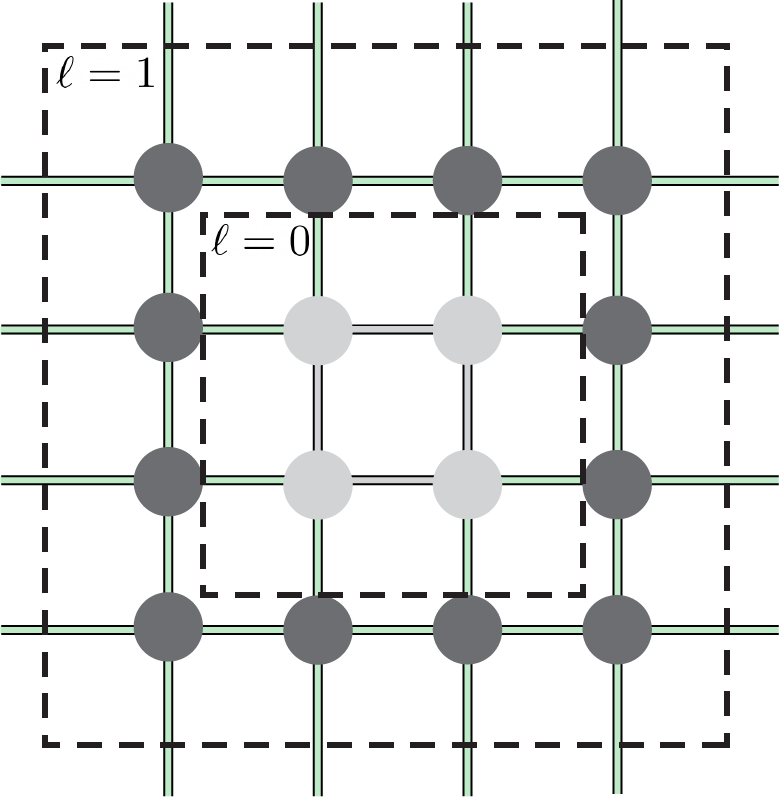}
    \caption{Schematic of fluid drainage proceeding from the initialization of a simulation. The light gray pore bodies connected by light gray pore throats represent the injection source used in all simulations. The pore throats in layer 0, denoted $\ell = 0$, are invaded to fill the first layer of pore bodies, shown in darker gray ($\ell = 1$). The volume eroded from the pore bodies in the first layer is then eroded into its connected pore throats, which are shown in mint green. We expect clogging to occur at $\ell = 1$, when the ratio of available pore throats to previously invaded pore bodies provides a value of $\alpha = 4/3$.}
    \label{a explanation}
\end{figure}

\begin{figure}[htp!]
    \centering
    \includegraphics[width = 0.5\textwidth]{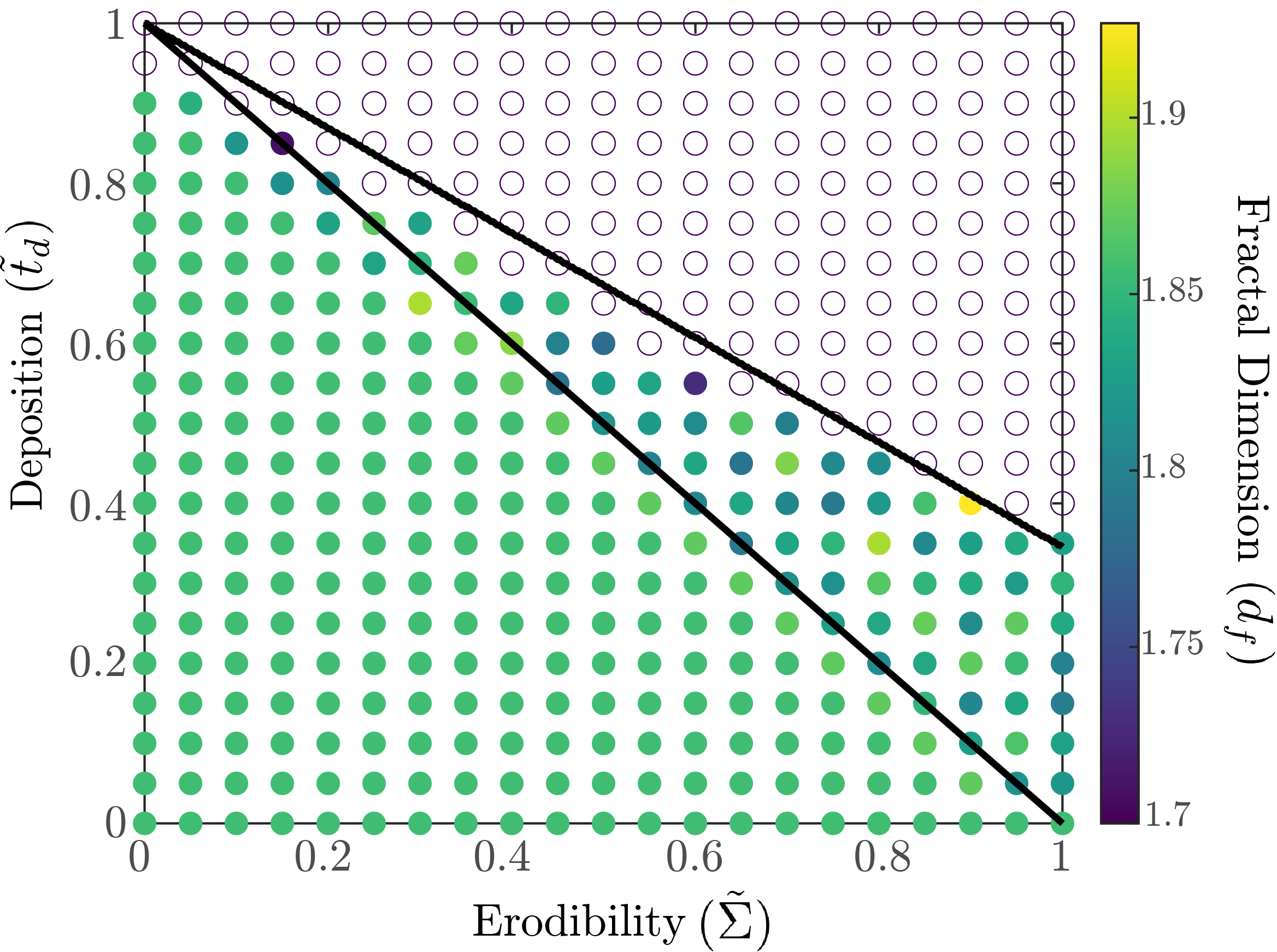}
    \caption{State diagram for the same simulations as in Fig. 3 of the main text, but instead showing  the fractal dimension $d_f$ of the nonwetting fluid displacement pathway after drainage through the medium has concluded. In the capillary fingering regime, our simulations return a constant fractal dimension, $d_f = 1.86 \pm 0.04$. However, in the erosion-enhanced fingering regime, the measured fractal dimensions are more varied, indicating that different $(\tilde{\Sigma}, \tilde{t}_d)$ combinations yield nonwetting fluid patterns with more varied ramification: we find a maximum measured fractal dimension of $d_f = 1.93 \pm 0.01$ for $(\tilde{\Sigma},\tilde{t}_d) = (0.9, 0.4)$, indicating a slightly more compact pathway, and a minimum measured fractal dimension of $d_f = 1.71 \pm 0.05$ for $(\tilde{\Sigma},\tilde{t}_d) = (0.15, 0.85)$, indicating a slightly more ramified pathway than capillary fingering. The empty circles indicate the rapid clogging regime, in which the filled volume fraction of the pore network is too low to accurately obtain $d_f$.}
    \label{fractal dimension}
\end{figure}

\begin{figure*}[h]
    \centering
    \includegraphics[width = \textwidth]{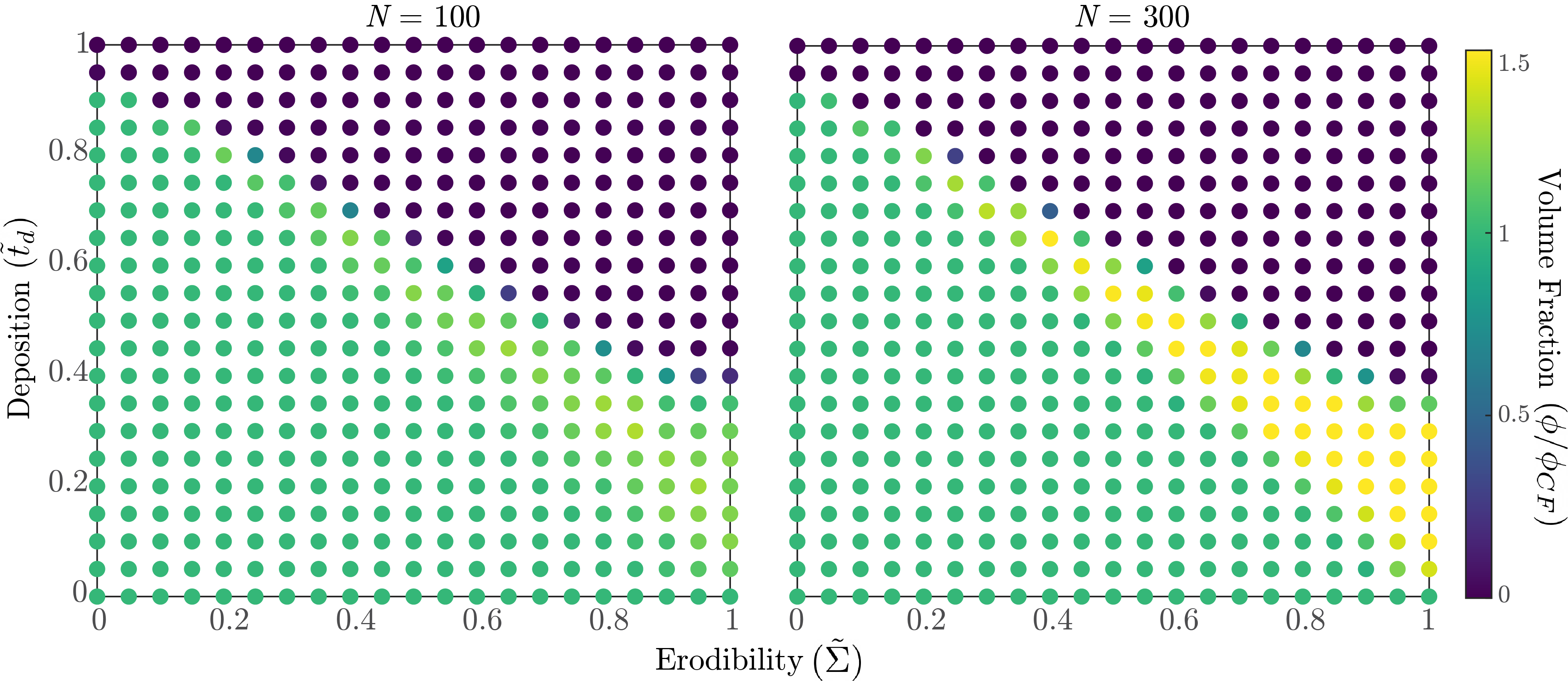}
    \caption{Additional state diagrams of nonwetting fluid filled volume fraction ($\phi/\phi_{CF}$) for networks with different $N = 100$ (left) and $N = 300$ (right), across the full range of deposition ($\tilde{t}_d$) and erodibility ($\tilde{\Sigma}$) values. For both system sizes, we again observe the emergence of capillary fingering, rapid clogging, and erosion enhanced fingering, with the boundaries between these different drainage behaviors remain unchanging and in good agreement with the results shown in the main text. The magnitude of $\phi/\phi_{CF}$ increases slightly, and then converges to $\phi/\phi_{CF} \approx 1.7$, with increasing $N$.}
    \label{N}
\end{figure*}

\begin{figure*}[h]
    \centering
    \includegraphics[width = \textwidth]{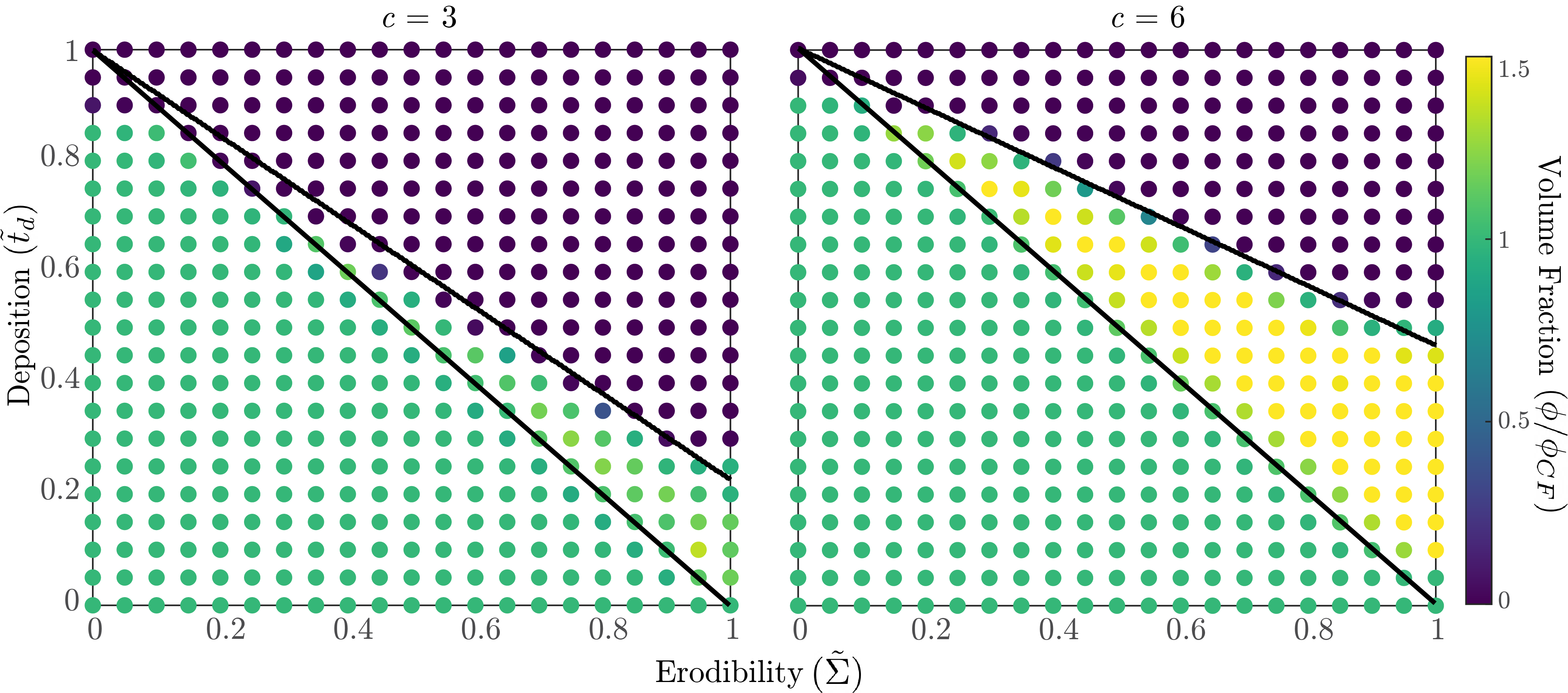}
    \caption{Additional state diagrams of nonwetting fluid filled volume fraction ($\phi/\phi_{CF}$) for networks with different connectivity, $c = 3$ (left) and $c = 6$ (right), across the full range of deposition ($\tilde{t}_d$) and erodibility ($\tilde{\Sigma}$) values. In both cases, we again observe the emergence of capillary fingering, rapid clogging, and erosion enhanced fingering, as in the main text. The erosion-enhanced fingering regime spans a smaller (larger) range of $(\tilde{t}_d,\tilde{\Sigma})$, and the corresponding $\phi/\phi_{CF}$ is smaller (larger), for the case of $c = 3$ ($c = 6$). These changes are captured by our theory when we account for network connectivities. When $c=3$, $\alpha = 2/3$, and when $c=6$, $\alpha = 5/2$.}
    \label{c}
\end{figure*}

\begin{figure*}[h]
    \centering
    \includegraphics[width = \textwidth]{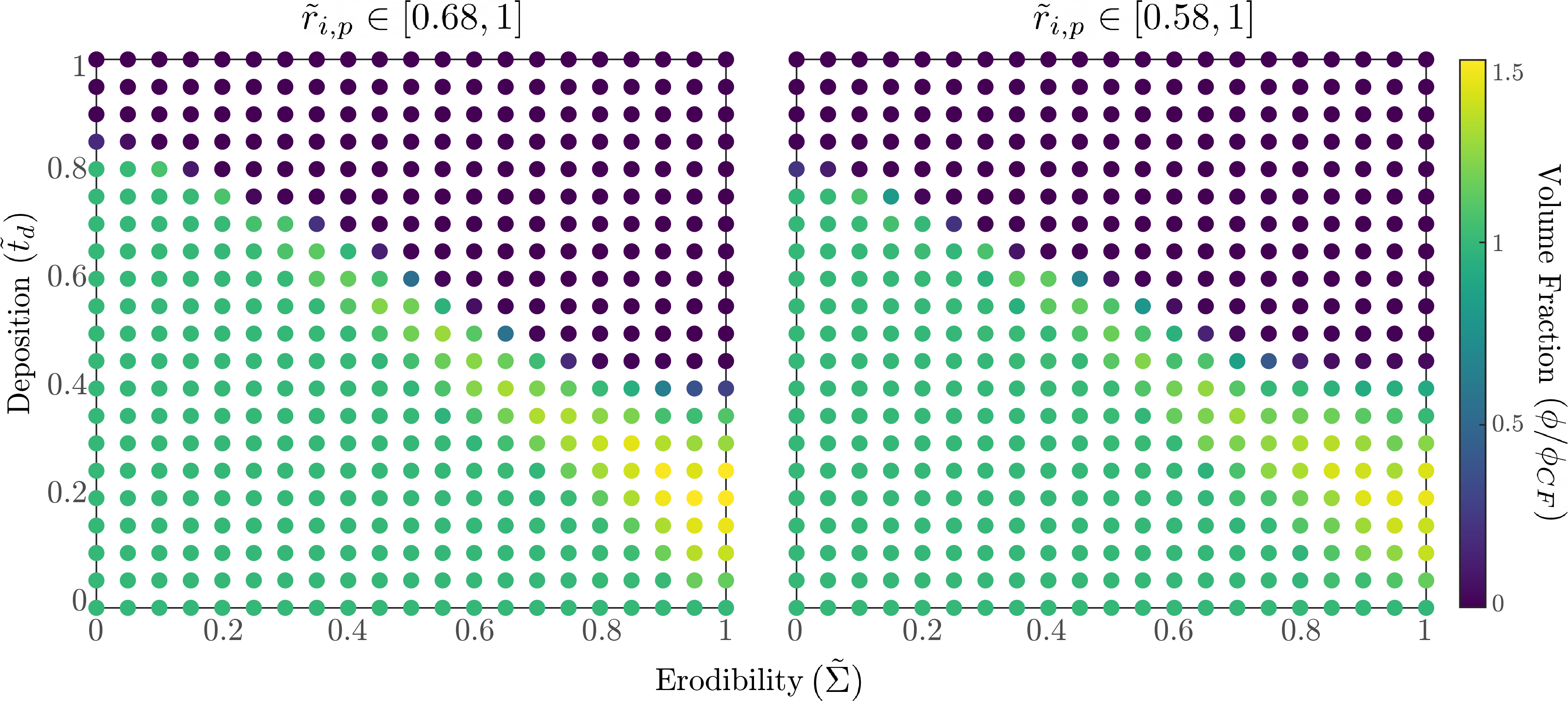}
    \caption{Additional state diagrams of nonwetting fluid filled volume fraction ($\phi/\phi_{CF}$) for wider uniform distributions, $\tilde{r}_{i,p}\in[0.68, 1]$ (left) and $\tilde{r}_{i,p}\in[0.58, 1]$ (right), across the full range of deposition ($\tilde{t}_d$) and erodibility ($\tilde{\Sigma}$) values. In both cases, we again observe the emergence of capillary fingering, rapid clogging, and erosion enhanced fingering, as in the main text. The erosion-enhanced fingering regime spans a smaller range of $(\tilde{t}_d,\tilde{\Sigma})$ as the distributions become wider.}
    \label{uniform distributions}
\end{figure*} 

\begin{figure*}
    \centering
    \includegraphics[width = \textwidth]{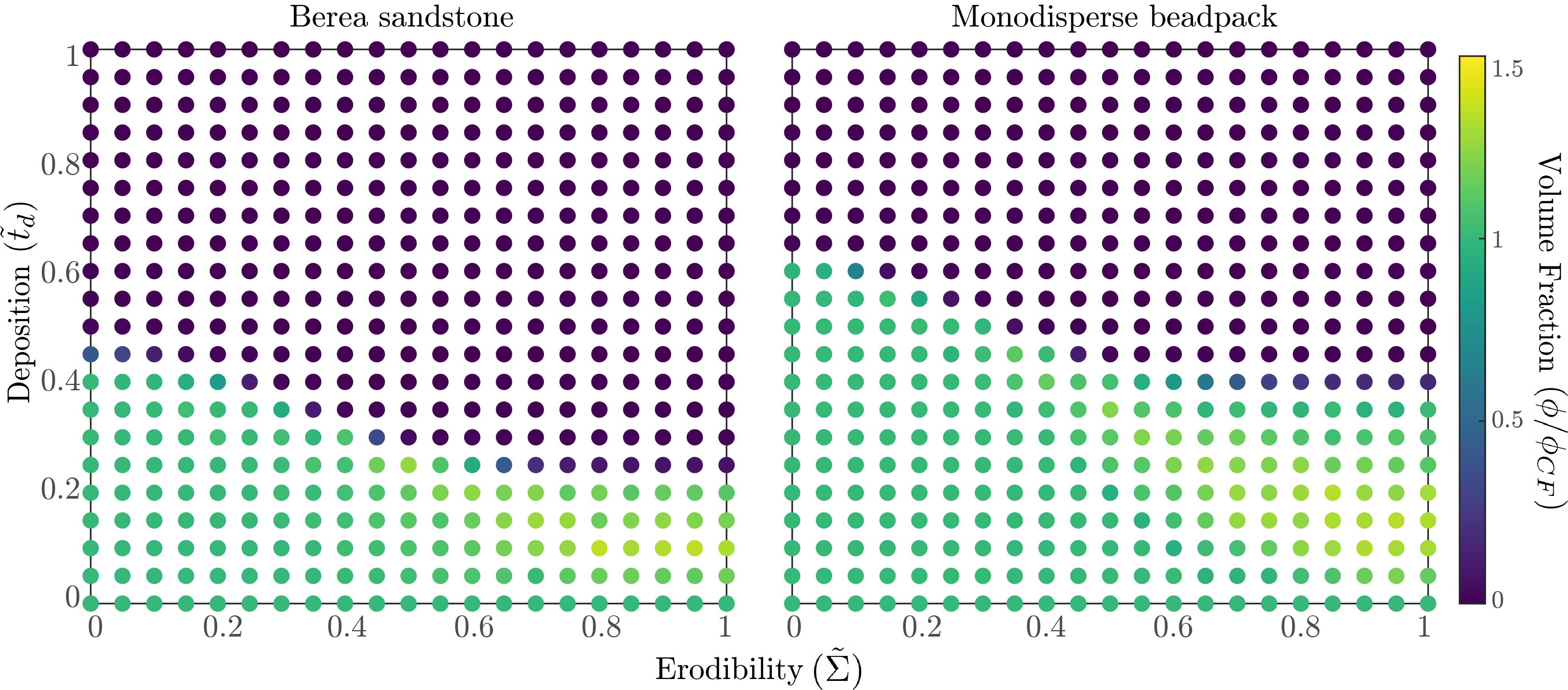}
    \caption{Additional state diagrams of nonwetting fluid filled volume fraction ($\phi/\phi_{CF}$) for non-uniform $\rho(r_{i,p})$ that are representative of two real-world examples as obtained from \cite{jerauld1990effect}: Berea sandstone (left) and a monodisperse bead packing (right). The Berea sandstone has $\rho(r_{i,p})=\frac{15}{4(r_{max}-r_{min})}\left(1-\frac{r_{i,p}-r_{min}}{r_{max}-r_{min}}\right)\sqrt{\frac{r_{i,p}-r_{min}}{r_{max}-r_{min}}}$, with $r_{min} = 1$ \textmu m and $r_{max} = 25$ \textmu m. The bead packing has $\rho(r_{i,p}) = \frac{6}{(r_{max}-r_{min})}\sqrt{\frac{r_{i,p}-r_{min}}{r_{max}-r_{min}}}\sqrt{1-\frac{r_{i,p}-r_{min}}{r_{max}-r_{min}}}$, with $r_{min} = 15$ \textmu m and $r_{max} = 40$ \textmu m. In both cases, we again observe the emergence of capillary fingering, rapid clogging, and erosion enhanced fingering, as in the main text.}    
    \label{jerauld and salter}
\end{figure*}

\subsection{Distribution of eroded material across adjacent connected pores}
To estimate how much material eroded from drained throat $i$ is redeposited into the non-drained throats $j$ that are directly connected to it, we consider the pressure drop $\delta p_j$ across each of $j$. Because the length of an individual pore throat, assumed to be uniform throughout the network, is much smaller than the overall length of the pore network (i.e., $N\gg1$), we assume that $\delta p_j$ is approximately constant across each downstream pore $j$, as given by the Hagen–Poiseuille equation. Thus, the flux of material into each throat $j$ is proportionate to $r_j^4$; we therefore assume that the new volume added to each of the $n$ connected throats $j$, $\delta V_j$, after a volume $V_{\rm{erode}}$ is eroded by drainage in throat $i$ is given by $\delta V_j = \frac{r_j^4}{\sum_{k = 1}^n r_k^4}V_{\rm{erode}}$. However, if $\delta V_j$ causes $\tilde{r}_{j}<0$, the excess volume is returned to the drained throat $i$ to conserve mass.

\subsection{Onset of rapid clogging}To estimate when pore clogging is so prevalent that it ``chokes off" fluid drainage, causing irreversible clogging, we balance the volume of solid material that can be eroded from a pore $i$, $V_{\rm{erode}}\propto\left(2\gamma/\sigma_y\right)^2-\left({r}_{i,p}-{t}^{*}_d\right)^2$, with the cumulative total available volume in the adjacent connected non-drained pore throats $j$, $V_{\rm{available}}\propto \left({r}_{j,p}-{t}^{*}_d\right)^2$. For tractability of computation, we make the assumption that both $r_{i,p}$ and $r_{j,p} \sim r_{p, \rm{max}}$. Thus $V_{\rm{erode}}\propto \tilde{\Sigma}^2-\left(1-\tilde{{t}}^{*}_d\right)^2$ and $V_{\rm{available}}\propto \left(1-\tilde{{t}}^{*}_d\right)^2$. This assumption that $r_{i,p}$ and $r_{j,p} \sim r_{p, \rm{max}}$ allows us to make the approximation that fluid drainage will expand radially in sequential annular ``layers'' from the central injection point (\figref{a explanation})---as opposed to the ramified invasion patterns typical of invasion percolation. On a square lattice of connectivity $c = 4$, layer $\ell$ experiences $8\ell+4$ invasions, yielding a total eroded volume $(8\ell+4)V_{erode}$, which gets redeposited onto $8(\ell+1)$ available pore throats with a total available volume of $8(\ell+1)V_{available}$. Thus, taking a mean-field approximation layer by layer, we expect clogging to occur at $\ell = 1$ when $(8\ell+4)V_{erode}\sim8(\ell+1)V_{available}$, or $V_{erode}\sim\frac{4}{3}V_{available}$. 

The same result can similarly be obtained for lattices with $c=3$ and $c=6$. For a lattice with connectivity $c=3$, layer $\ell$ experiences $6(2\ell+1)$ invasions, yielding a total eroded volume $6(2\ell+1)V_{erode}$, which gets redeposited onto $6(\ell+1)$ available pore throats with a total available volume of $6(\ell+1)V_{available}$. If we similarly expect clogging to occur at $\ell = 1$, $V_{erode}\sim\frac{2}{3}V_{available}$ (\figref{c}, left). For a lattice with connectivity $c=6$, layer $\ell$ experiences $6\ell+6$ invasions, yielding a total eroded volume $(6\ell+6)V_{erode}$, which gets redeposited onto $12(\ell+1)+6$ available pore throats with a total available volume of $(12(\ell+1)+6)V_{available}$. If we similarly expect clogging to occur at $\ell=1$, $V_{erode}\sim\frac{5}{2}V_{available}$ (\figref{c}, right).

\subsection{Captions for Supplementary Videos}
\noindent \textbf{Movie S1.} Invasion percolation in a pore network with $(\tilde{\Sigma}, \tilde{t}_d) = (0, 0)$ shows the traditional invasion percolation algorithm and an example of a classic capillary fingering pattern. Gray circles denote pore throats invaded by the nonwetting fluid and gray $+$ symbols denote invaded pore bodies. The left panel shows a view of the entire network. The blue box denotes the magnified view shown in the right panel for clarity.\\

\noindent \textbf{Movie S2.} Invasion percolation in a pore network with $(\tilde{\Sigma}, \tilde{t}_d) = (0, 0.25)$ returns the traditional invasion percolation algorithm and shows an example of a classic capillary fingering pattern. Gray circles denote pore throats invaded by the nonwetting fluid and gray $+$ symbols denote invaded pore bodies. The left panel shows a view of the entire network. The blue box denotes the magnified view shown in the right panel for clarity.\\

\noindent \textbf{Movie S3.} Invasion percolation in a pore network with $(\tilde{\Sigma}, \tilde{t}_d) = (0, 0.95)$ shows that only a few invasions occur before clogging occurs, choking off subsequent flow. Gray circles denote pore throats invaded by the nonwetting fluid and gray $+$ symbols denote invaded pore bodies. Purple circles denote clogged pore throats. The left panel shows a view of the entire network. The blue box denotes the magnified view shown in the right panel for clarity.\\

\noindent \textbf{Movie S4.} Invasion percolation in a pore network with $(\tilde{\Sigma}, \tilde{t}_d) = (0.8, 0.25)$ shows a markedly different nonwetting fluid invasion pattern that fills more of the pore space than capillary fingering alone. No clogging occurs during this simulation. Gray circles denote pore throats invaded by the nonwetting fluid and gray $+$ symbols denote invaded pore bodies. The left panel shows a view of the entire network. The blue box denotes the magnified view shown in the right panel for clarity.\\

\noindent \textbf{Movie S5.} Invasion percolation in a pore network with $(\tilde{\Sigma}, \tilde{t}_d) = (0.8, 0.35)$ shows another nonwetting fluid invasion pattern that fills more of the pore space than capillary fingering alone. Intermittent clogging also occurs in this simulation. Gray circles denote pore throats invaded by the nonwetting fluid and gray $+$ symbols denote invaded pore bodies. Purple circles denote clogged pore throats. The left panel shows a view of the entire network. The blue box denotes the magnified view shown in the right panel for clarity.\\

\noindent \textbf{Movie S6.} Invasion percolation in a pore network with $(\tilde{\Sigma}, \tilde{t}_d) = (0.8, 0.45)$ shows another nonwetting fluid invasion pattern that appears dense, but clogging chockes off flow before the nonwetting fluid can percolate through the network. Gray circles denote pore throats invaded by the nonwetting fluid and gray $+$ symbols denote invaded pore bodies. Purple circles denote clogged pore throats. The left panel shows a view of the entire network. The blue box denotes the magnified view shown in the right panel for clarity.


\begin{thebibliography}{73}%
\makeatletter
\providecommand \@ifxundefined [1]{%
 \@ifx{#1\undefined}
}%
\providecommand \@ifnum [1]{%
 \ifnum #1\expandafter \@firstoftwo
 \else \expandafter \@secondoftwo
 \fi
}%
\providecommand \@ifx [1]{%
 \ifx #1\expandafter \@firstoftwo
 \else \expandafter \@secondoftwo
 \fi
}%
\providecommand \natexlab [1]{#1}%
\providecommand \enquote  [1]{``#1''}%
\providecommand \bibnamefont  [1]{#1}%
\providecommand \bibfnamefont [1]{#1}%
\providecommand \citenamefont [1]{#1}%
\providecommand \href@noop [0]{\@secondoftwo}%
\providecommand \href [0]{\begingroup \@sanitize@url \@href}%
\providecommand \@href[1]{\@@startlink{#1}\@@href}%
\providecommand \@@href[1]{\endgroup#1\@@endlink}%
\providecommand \@sanitize@url [0]{\catcode `\\12\catcode `\$12\catcode
  `\&12\catcode `\#12\catcode `\^12\catcode `\_12\catcode `\%12\relax}%
\providecommand \@@startlink[1]{}%
\providecommand \@@endlink[0]{}%
\providecommand \url  [0]{\begingroup\@sanitize@url \@url }%
\providecommand \@url [1]{\endgroup\@href {#1}{\urlprefix }}%
\providecommand \urlprefix  [0]{URL }%
\providecommand \Eprint [0]{\href }%
\providecommand \doibase [0]{https://doi.org/}%
\providecommand \selectlanguage [0]{\@gobble}%
\providecommand \bibinfo  [0]{\@secondoftwo}%
\providecommand \bibfield  [0]{\@secondoftwo}%
\providecommand \translation [1]{[#1]}%
\providecommand \BibitemOpen [0]{}%
\providecommand \bibitemStop [0]{}%
\providecommand \bibitemNoStop [0]{.\EOS\space}%
\providecommand \EOS [0]{\spacefactor3000\relax}%
\providecommand \BibitemShut  [1]{\csname bibitem#1\endcsname}%
\let\auto@bib@innerbib\@empty
\bibitem [{\citenamefont {Bethke}\ \emph {et~al.}(1991)\citenamefont {Bethke},
  \citenamefont {Reed},\ and\ \citenamefont {Oltz}}]{bethke1991long}%
  \BibitemOpen
  \bibfield  {author} {\bibinfo {author} {\bibfnamefont {C.~M.}\ \bibnamefont
  {Bethke}}, \bibinfo {author} {\bibfnamefont {J.~D.}\ \bibnamefont {Reed}},\
  and\ \bibinfo {author} {\bibfnamefont {D.~F.}\ \bibnamefont {Oltz}},\
  }\bibfield  {title} {\bibinfo {title} {Long-range petroleum migration in the
  illinois basin},\ }\href@noop {} {\bibfield  {journal} {\bibinfo  {journal}
  {AAPG Bulletin}\ }\textbf {\bibinfo {volume} {75}},\ \bibinfo {pages} {925}
  (\bibinfo {year} {1991})}\BibitemShut {NoStop}%
\bibitem [{\citenamefont {Kueper}\ and\ \citenamefont
  {Frind}(1991)}]{kueper1991two}%
  \BibitemOpen
  \bibfield  {author} {\bibinfo {author} {\bibfnamefont {B.~H.}\ \bibnamefont
  {Kueper}}\ and\ \bibinfo {author} {\bibfnamefont {E.~O.}\ \bibnamefont
  {Frind}},\ }\bibfield  {title} {\bibinfo {title} {Two-phase flow in
  heterogeneous porous media: 1. model development},\ }\href@noop {} {\bibfield
   {journal} {\bibinfo  {journal} {Water Resources Research}\ }\textbf
  {\bibinfo {volume} {27}},\ \bibinfo {pages} {1049} (\bibinfo {year}
  {1991})}\BibitemShut {NoStop}%
\bibitem [{\citenamefont {Dawson}\ and\ \citenamefont
  {Roberts}(1997)}]{dawson1997influence}%
  \BibitemOpen
  \bibfield  {author} {\bibinfo {author} {\bibfnamefont {H.~E.}\ \bibnamefont
  {Dawson}}\ and\ \bibinfo {author} {\bibfnamefont {P.~V.}\ \bibnamefont
  {Roberts}},\ }\bibfield  {title} {\bibinfo {title} {Influence of viscous,
  gravitational, and capillary forces on {DNAPL} saturation},\ }\href@noop {}
  {\bibfield  {journal} {\bibinfo  {journal} {Groundwater}\ }\textbf {\bibinfo
  {volume} {35}},\ \bibinfo {pages} {261} (\bibinfo {year} {1997})}\BibitemShut
  {NoStop}%
\bibitem [{\citenamefont {Levy}\ \emph {et~al.}(2003)\citenamefont {Levy},
  \citenamefont {Culligan},\ and\ \citenamefont
  {Germaine}}]{levy2003modelling}%
  \BibitemOpen
  \bibfield  {author} {\bibinfo {author} {\bibfnamefont {L.~C.}\ \bibnamefont
  {Levy}}, \bibinfo {author} {\bibfnamefont {P.~J.}\ \bibnamefont {Culligan}},\
  and\ \bibinfo {author} {\bibfnamefont {J.~T.}\ \bibnamefont {Germaine}},\
  }\bibfield  {title} {\bibinfo {title} {Modelling of {DNAPL} behavior in
  vertical fractures},\ }\href@noop {} {\bibfield  {journal} {\bibinfo
  {journal} {International Journal of Physical Modelling in Geotechnics}\
  }\textbf {\bibinfo {volume} {3}},\ \bibinfo {pages} {01} (\bibinfo {year}
  {2003})}\BibitemShut {NoStop}%
\bibitem [{\citenamefont {Dandekar}(2006)}]{dandekar2006petroleum}%
  \BibitemOpen
  \bibfield  {author} {\bibinfo {author} {\bibfnamefont {A.~Y.}\ \bibnamefont
  {Dandekar}},\ }\href@noop {} {\emph {\bibinfo {title} {Petroleum reservoir
  rock and fluid properties}}}\ (\bibinfo  {publisher} {CRC Press},\ \bibinfo
  {year} {2006})\BibitemShut {NoStop}%
\bibitem [{\citenamefont {Benson}\ and\ \citenamefont
  {Orr}(2008)}]{benson2008carbon}%
  \BibitemOpen
  \bibfield  {author} {\bibinfo {author} {\bibfnamefont {S.~M.}\ \bibnamefont
  {Benson}}\ and\ \bibinfo {author} {\bibfnamefont {F.~M.}\ \bibnamefont
  {Orr}},\ }\bibfield  {title} {\bibinfo {title} {Carbon dioxide capture and
  storage},\ }\href@noop {} {\bibfield  {journal} {\bibinfo  {journal} {MRS
  Bulletin}\ }\textbf {\bibinfo {volume} {33}},\ \bibinfo {pages} {303}
  (\bibinfo {year} {2008})}\BibitemShut {NoStop}%
\bibitem [{\citenamefont {Cueto-Felgueroso}\ and\ \citenamefont
  {Juanes}(2008)}]{cueto2008nonlocal}%
  \BibitemOpen
  \bibfield  {author} {\bibinfo {author} {\bibfnamefont {L.}~\bibnamefont
  {Cueto-Felgueroso}}\ and\ \bibinfo {author} {\bibfnamefont {R.}~\bibnamefont
  {Juanes}},\ }\bibfield  {title} {\bibinfo {title} {Nonlocal interface
  dynamics and pattern formation in gravity-driven unsaturated flow through
  porous media},\ }\href@noop {} {\bibfield  {journal} {\bibinfo  {journal}
  {Physical Review Letters}\ }\textbf {\bibinfo {volume} {101}},\ \bibinfo
  {pages} {244504} (\bibinfo {year} {2008})}\BibitemShut {NoStop}%
\bibitem [{\citenamefont {Neufeld}\ and\ \citenamefont
  {Huppert}(2009)}]{neufeld2009modelling}%
  \BibitemOpen
  \bibfield  {author} {\bibinfo {author} {\bibfnamefont {J.~A.}\ \bibnamefont
  {Neufeld}}\ and\ \bibinfo {author} {\bibfnamefont {H.~E.}\ \bibnamefont
  {Huppert}},\ }\bibfield  {title} {\bibinfo {title} {Modelling carbon dioxide
  sequestration in layered strata},\ }\href@noop {} {\bibfield  {journal}
  {\bibinfo  {journal} {Journal of Fluid Mechanics}\ }\textbf {\bibinfo
  {volume} {625}},\ \bibinfo {pages} {353} (\bibinfo {year}
  {2009})}\BibitemShut {NoStop}%
\bibitem [{\citenamefont {Bear}\ and\ \citenamefont
  {Cheng}(2010)}]{bear2010modeling}%
  \BibitemOpen
  \bibfield  {author} {\bibinfo {author} {\bibfnamefont {J.}~\bibnamefont
  {Bear}}\ and\ \bibinfo {author} {\bibfnamefont {A.~H.-D.}\ \bibnamefont
  {Cheng}},\ }\href@noop {} {\emph {\bibinfo {title} {Modeling groundwater flow
  and contaminant transport}}},\ Vol.~\bibinfo {volume} {23}\ (\bibinfo
  {publisher} {Springer},\ \bibinfo {year} {2010})\BibitemShut {NoStop}%
\bibitem [{\citenamefont {MacMinn}\ \emph {et~al.}(2010)\citenamefont
  {MacMinn}, \citenamefont {Szulczewski},\ and\ \citenamefont
  {Juanes}}]{macminn2010co2}%
  \BibitemOpen
  \bibfield  {author} {\bibinfo {author} {\bibfnamefont {C.~W.}\ \bibnamefont
  {MacMinn}}, \bibinfo {author} {\bibfnamefont {M.~L.}\ \bibnamefont
  {Szulczewski}},\ and\ \bibinfo {author} {\bibfnamefont {R.}~\bibnamefont
  {Juanes}},\ }\bibfield  {title} {\bibinfo {title} {{CO$_2$} migration in
  saline aquifers. part 1. {Capillary} trapping under slope and groundwater
  flow},\ }\href@noop {} {\bibfield  {journal} {\bibinfo  {journal} {Journal of
  Fluid Mechanics}\ }\textbf {\bibinfo {volume} {662}},\ \bibinfo {pages} {329}
  (\bibinfo {year} {2010})}\BibitemShut {NoStop}%
\bibitem [{\citenamefont {Saadatpoor}\ \emph {et~al.}(2010)\citenamefont
  {Saadatpoor}, \citenamefont {Bryant},\ and\ \citenamefont
  {Sepehrnoori}}]{saadatpoor2010new}%
  \BibitemOpen
  \bibfield  {author} {\bibinfo {author} {\bibfnamefont {E.}~\bibnamefont
  {Saadatpoor}}, \bibinfo {author} {\bibfnamefont {S.~L.}\ \bibnamefont
  {Bryant}},\ and\ \bibinfo {author} {\bibfnamefont {K.}~\bibnamefont
  {Sepehrnoori}},\ }\bibfield  {title} {\bibinfo {title} {New trapping
  mechanism in carbon sequestration},\ }\href@noop {} {\bibfield  {journal}
  {\bibinfo  {journal} {Transport in Porous Media}\ }\textbf {\bibinfo {volume}
  {82}},\ \bibinfo {pages} {3} (\bibinfo {year} {2010})}\BibitemShut {NoStop}%
\bibitem [{\citenamefont {Bandara}\ \emph {et~al.}(2011)\citenamefont
  {Bandara}, \citenamefont {Tartakovsky},\ and\ \citenamefont
  {Palmer}}]{bandara2011pore}%
  \BibitemOpen
  \bibfield  {author} {\bibinfo {author} {\bibfnamefont {U.~C.}\ \bibnamefont
  {Bandara}}, \bibinfo {author} {\bibfnamefont {A.~M.}\ \bibnamefont
  {Tartakovsky}},\ and\ \bibinfo {author} {\bibfnamefont {B.~J.}\ \bibnamefont
  {Palmer}},\ }\bibfield  {title} {\bibinfo {title} {Pore-scale study of
  capillary trapping mechanism during {CO$_2$} injection in geological
  formations},\ }\href@noop {} {\bibfield  {journal} {\bibinfo  {journal}
  {International Journal of Greenhouse Gas Control}\ }\textbf {\bibinfo
  {volume} {5}},\ \bibinfo {pages} {1566} (\bibinfo {year} {2011})}\BibitemShut
  {NoStop}%
\bibitem [{\citenamefont {Sahimi}(2011)}]{sahimi2011flow}%
  \BibitemOpen
  \bibfield  {author} {\bibinfo {author} {\bibfnamefont {M.}~\bibnamefont
  {Sahimi}},\ }\href@noop {} {\emph {\bibinfo {title} {Flow and transport in
  porous media and fractured rock: from classical methods to modern
  approaches}}}\ (\bibinfo  {publisher} {John Wiley \& Sons},\ \bibinfo {year}
  {2011})\BibitemShut {NoStop}%
\bibitem [{\citenamefont {Berg}\ and\ \citenamefont
  {Ott}(2012)}]{berg2012stability}%
  \BibitemOpen
  \bibfield  {author} {\bibinfo {author} {\bibfnamefont {S.}~\bibnamefont
  {Berg}}\ and\ \bibinfo {author} {\bibfnamefont {H.}~\bibnamefont {Ott}},\
  }\bibfield  {title} {\bibinfo {title} {Stability of {CO$_2$}--brine
  immiscible displacement},\ }\href@noop {} {\bibfield  {journal} {\bibinfo
  {journal} {International Journal of Greenhouse Gas Control}\ }\textbf
  {\bibinfo {volume} {11}},\ \bibinfo {pages} {188} (\bibinfo {year}
  {2012})}\BibitemShut {NoStop}%
\bibitem [{\citenamefont {Carmo}\ \emph {et~al.}(2013)\citenamefont {Carmo},
  \citenamefont {Fritz}, \citenamefont {Mergel},\ and\ \citenamefont
  {Stolten}}]{carmo2013comprehensive}%
  \BibitemOpen
  \bibfield  {author} {\bibinfo {author} {\bibfnamefont {M.}~\bibnamefont
  {Carmo}}, \bibinfo {author} {\bibfnamefont {D.~L.}\ \bibnamefont {Fritz}},
  \bibinfo {author} {\bibfnamefont {J.}~\bibnamefont {Mergel}},\ and\ \bibinfo
  {author} {\bibfnamefont {D.}~\bibnamefont {Stolten}},\ }\bibfield  {title}
  {\bibinfo {title} {A comprehensive review on {PEM} water electrolysis},\
  }\href@noop {} {\bibfield  {journal} {\bibinfo  {journal} {International
  Journal of Hydrogen Energy}\ }\textbf {\bibinfo {volume} {38}},\ \bibinfo
  {pages} {4901} (\bibinfo {year} {2013})}\BibitemShut {NoStop}%
\bibitem [{\citenamefont {Lee}\ \emph {et~al.}(2016)\citenamefont {Lee},
  \citenamefont {Banerjee}, \citenamefont {Arbabi}, \citenamefont {Hinebaugh},\
  and\ \citenamefont {Bazylak}}]{lee2016porous}%
  \BibitemOpen
  \bibfield  {author} {\bibinfo {author} {\bibfnamefont {C.~H.}\ \bibnamefont
  {Lee}}, \bibinfo {author} {\bibfnamefont {R.}~\bibnamefont {Banerjee}},
  \bibinfo {author} {\bibfnamefont {F.}~\bibnamefont {Arbabi}}, \bibinfo
  {author} {\bibfnamefont {J.}~\bibnamefont {Hinebaugh}},\ and\ \bibinfo
  {author} {\bibfnamefont {A.}~\bibnamefont {Bazylak}},\ }\bibfield  {title}
  {\bibinfo {title} {Porous transport layer related mass transport losses in
  polymer electrolyte membrane electrolysis: A review},\ }in\ \href@noop {}
  {\emph {\bibinfo {booktitle} {International Conference on Nanochannels,
  Microchannels, and Minichannels}}},\ Vol.\ \bibinfo {volume} {50343}\
  (\bibinfo {organization} {American Society of Mechanical Engineers},\
  \bibinfo {year} {2016})\ p.\ \bibinfo {pages} {V001T07A003}\BibitemShut
  {NoStop}%
\bibitem [{\citenamefont {Bazyar}\ \emph {et~al.}(2018)\citenamefont {Bazyar},
  \citenamefont {Lv}, \citenamefont {Wood}, \citenamefont {Porada},
  \citenamefont {Lohse},\ and\ \citenamefont {Lammertink}}]{bazyar2018liquid}%
  \BibitemOpen
  \bibfield  {author} {\bibinfo {author} {\bibfnamefont {H.}~\bibnamefont
  {Bazyar}}, \bibinfo {author} {\bibfnamefont {P.}~\bibnamefont {Lv}}, \bibinfo
  {author} {\bibfnamefont {J.~A.}\ \bibnamefont {Wood}}, \bibinfo {author}
  {\bibfnamefont {S.}~\bibnamefont {Porada}}, \bibinfo {author} {\bibfnamefont
  {D.}~\bibnamefont {Lohse}},\ and\ \bibinfo {author} {\bibfnamefont {R.~G.}\
  \bibnamefont {Lammertink}},\ }\bibfield  {title} {\bibinfo {title}
  {Liquid--liquid displacement in slippery liquid-infused membranes (slims)},\
  }\href@noop {} {\bibfield  {journal} {\bibinfo  {journal} {Soft Matter}\
  }\textbf {\bibinfo {volume} {14}},\ \bibinfo {pages} {1780} (\bibinfo {year}
  {2018})}\BibitemShut {NoStop}%
\bibitem [{\citenamefont {Knackstedt}\ and\ \citenamefont
  {Paterson}(2009)}]{knackstedt2009invasion}%
  \BibitemOpen
  \bibfield  {author} {\bibinfo {author} {\bibfnamefont {M.}~\bibnamefont
  {Knackstedt}}\ and\ \bibinfo {author} {\bibfnamefont {L.}~\bibnamefont
  {Paterson}},\ }\bibinfo {title} {Invasion percolation},\ in\ \href
  {https://doi.org/10.1007/978-0-387-30440-3_294} {\emph {\bibinfo {booktitle}
  {Encyclopedia of Complexity and Systems Science}}},\ \bibinfo {editor}
  {edited by\ \bibinfo {editor} {\bibfnamefont {R.~A.}\ \bibnamefont
  {Meyers}}}\ (\bibinfo  {publisher} {Springer New York},\ \bibinfo {address}
  {New York, NY},\ \bibinfo {year} {2009})\ pp.\ \bibinfo {pages}
  {4947--4960}\BibitemShut {NoStop}%
\bibitem [{\citenamefont {Blunt}(2017)}]{blunt2017multiphase}%
  \BibitemOpen
  \bibfield  {author} {\bibinfo {author} {\bibfnamefont {M.~J.}\ \bibnamefont
  {Blunt}},\ }\href@noop {} {\emph {\bibinfo {title} {Multiphase flow in
  permeable media: A pore-scale perspective}}}\ (\bibinfo  {publisher}
  {Cambridge University Press},\ \bibinfo {year} {2017})\BibitemShut {NoStop}%
\bibitem [{\citenamefont {Wilkinson}\ and\ \citenamefont
  {Willemsen}(1983)}]{wilkinson1983invasion}%
  \BibitemOpen
  \bibfield  {author} {\bibinfo {author} {\bibfnamefont {D.}~\bibnamefont
  {Wilkinson}}\ and\ \bibinfo {author} {\bibfnamefont {J.~F.}\ \bibnamefont
  {Willemsen}},\ }\bibfield  {title} {\bibinfo {title} {Invasion percolation: a
  new form of percolation theory},\ }\href@noop {} {\bibfield  {journal}
  {\bibinfo  {journal} {Journal of Physics A: Mathematical and General}\
  }\textbf {\bibinfo {volume} {16}},\ \bibinfo {pages} {3365} (\bibinfo {year}
  {1983})}\BibitemShut {NoStop}%
\bibitem [{\citenamefont {Mayer}\ and\ \citenamefont
  {Stowe}(1965)}]{mayer1965mercury}%
  \BibitemOpen
  \bibfield  {author} {\bibinfo {author} {\bibfnamefont {R.~P.}\ \bibnamefont
  {Mayer}}\ and\ \bibinfo {author} {\bibfnamefont {R.~A.}\ \bibnamefont
  {Stowe}},\ }\bibfield  {title} {\bibinfo {title} {Mercury
  porosimetry—breakthrough pressure for penetration between packed spheres},\
  }\href@noop {} {\bibfield  {journal} {\bibinfo  {journal} {Journal of Colloid
  Science}\ }\textbf {\bibinfo {volume} {20}},\ \bibinfo {pages} {893}
  (\bibinfo {year} {1965})}\BibitemShut {NoStop}%
\bibitem [{\citenamefont {Lenormand}\ \emph {et~al.}(1983)\citenamefont
  {Lenormand}, \citenamefont {Zarcone},\ and\ \citenamefont
  {Sarr}}]{lenormand1983mechanisms}%
  \BibitemOpen
  \bibfield  {author} {\bibinfo {author} {\bibfnamefont {R.}~\bibnamefont
  {Lenormand}}, \bibinfo {author} {\bibfnamefont {C.}~\bibnamefont {Zarcone}},\
  and\ \bibinfo {author} {\bibfnamefont {A.}~\bibnamefont {Sarr}},\ }\bibfield
  {title} {\bibinfo {title} {Mechanisms of the displacement of one fluid by
  another in a network of capillary ducts},\ }\href@noop {} {\bibfield
  {journal} {\bibinfo  {journal} {Journal of Fluid Mechanics}\ }\textbf
  {\bibinfo {volume} {135}},\ \bibinfo {pages} {337} (\bibinfo {year}
  {1983})}\BibitemShut {NoStop}%
\bibitem [{\citenamefont {Lenormand}\ and\ \citenamefont
  {Zarcone}(1985)}]{lenormand1985invasion}%
  \BibitemOpen
  \bibfield  {author} {\bibinfo {author} {\bibfnamefont {R.}~\bibnamefont
  {Lenormand}}\ and\ \bibinfo {author} {\bibfnamefont {C.}~\bibnamefont
  {Zarcone}},\ }\bibfield  {title} {\bibinfo {title} {Invasion percolation in
  an etched network: measurement of a fractal dimension},\ }\href@noop {}
  {\bibfield  {journal} {\bibinfo  {journal} {Physical Review Letters}\
  }\textbf {\bibinfo {volume} {54}},\ \bibinfo {pages} {2226} (\bibinfo {year}
  {1985})}\BibitemShut {NoStop}%
\bibitem [{\citenamefont {Mason}\ and\ \citenamefont
  {Morrow}(1986)}]{mason1986meniscus}%
  \BibitemOpen
  \bibfield  {author} {\bibinfo {author} {\bibfnamefont {G.}~\bibnamefont
  {Mason}}\ and\ \bibinfo {author} {\bibfnamefont {N.}~\bibnamefont {Morrow}},\
  }\bibfield  {title} {\bibinfo {title} {Meniscus displacement curvatures of a
  perfectly wetting liquid in capillary pore throats formed by spheres},\
  }\href@noop {} {\bibfield  {journal} {\bibinfo  {journal} {Journal of Colloid
  and Interface Science}\ }\textbf {\bibinfo {volume} {109}},\ \bibinfo {pages}
  {46} (\bibinfo {year} {1986})}\BibitemShut {NoStop}%
\bibitem [{\citenamefont {Lenormand}\ and\ \citenamefont
  {Zarcone}(1989)}]{lenormand1989capillary}%
  \BibitemOpen
  \bibfield  {author} {\bibinfo {author} {\bibfnamefont {R.}~\bibnamefont
  {Lenormand}}\ and\ \bibinfo {author} {\bibfnamefont {C.}~\bibnamefont
  {Zarcone}},\ }\bibfield  {title} {\bibinfo {title} {Capillary fingering:
  percolation and fractal dimension},\ }\href@noop {} {\bibfield  {journal}
  {\bibinfo  {journal} {Transport in Porous Media}\ }\textbf {\bibinfo {volume}
  {4}},\ \bibinfo {pages} {599} (\bibinfo {year} {1989})}\BibitemShut {NoStop}%
\bibitem [{\citenamefont {Martys}\ \emph {et~al.}(1991)\citenamefont {Martys},
  \citenamefont {Cieplak},\ and\ \citenamefont {Robbins}}]{martys1991critical}%
  \BibitemOpen
  \bibfield  {author} {\bibinfo {author} {\bibfnamefont {N.}~\bibnamefont
  {Martys}}, \bibinfo {author} {\bibfnamefont {M.}~\bibnamefont {Cieplak}},\
  and\ \bibinfo {author} {\bibfnamefont {M.~O.}\ \bibnamefont {Robbins}},\
  }\bibfield  {title} {\bibinfo {title} {Critical phenomena in fluid invasion
  of porous media},\ }\href@noop {} {\bibfield  {journal} {\bibinfo  {journal}
  {Physical Review Letters}\ }\textbf {\bibinfo {volume} {66}},\ \bibinfo
  {pages} {1058} (\bibinfo {year} {1991})}\BibitemShut {NoStop}%
\bibitem [{\citenamefont {M{\aa}l{\o}y}\ \emph {et~al.}(1992)\citenamefont
  {M{\aa}l{\o}y}, \citenamefont {Furuberg}, \citenamefont {Feder},\ and\
  \citenamefont {J{\o}ssang}}]{maaloy1992dynamics}%
  \BibitemOpen
  \bibfield  {author} {\bibinfo {author} {\bibfnamefont {K.~J.}\ \bibnamefont
  {M{\aa}l{\o}y}}, \bibinfo {author} {\bibfnamefont {L.}~\bibnamefont
  {Furuberg}}, \bibinfo {author} {\bibfnamefont {J.}~\bibnamefont {Feder}},\
  and\ \bibinfo {author} {\bibfnamefont {T.}~\bibnamefont {J{\o}ssang}},\
  }\bibfield  {title} {\bibinfo {title} {Dynamics of slow drainage in porous
  media},\ }\href@noop {} {\bibfield  {journal} {\bibinfo  {journal} {Physical
  Review Letters}\ }\textbf {\bibinfo {volume} {68}},\ \bibinfo {pages} {2161}
  (\bibinfo {year} {1992})}\BibitemShut {NoStop}%
\bibitem [{\citenamefont {Toledo}\ \emph {et~al.}(1994)\citenamefont {Toledo},
  \citenamefont {Scriven},\ and\ \citenamefont {Davis}}]{toledo1994pore}%
  \BibitemOpen
  \bibfield  {author} {\bibinfo {author} {\bibfnamefont {P.~G.}\ \bibnamefont
  {Toledo}}, \bibinfo {author} {\bibfnamefont {L.}~\bibnamefont {Scriven}},\
  and\ \bibinfo {author} {\bibfnamefont {H.~T.}\ \bibnamefont {Davis}},\
  }\bibfield  {title} {\bibinfo {title} {Pore-space statistics and capillary
  pressure curves from volume-controlled porosimetry},\ }\href@noop {}
  {\bibfield  {journal} {\bibinfo  {journal} {SPE Formation Evaluation}\
  }\textbf {\bibinfo {volume} {9}},\ \bibinfo {pages} {46} (\bibinfo {year}
  {1994})}\BibitemShut {NoStop}%
\bibitem [{\citenamefont {Xu}\ \emph {et~al.}(1998)\citenamefont {Xu},
  \citenamefont {Yortsos},\ and\ \citenamefont {Salin}}]{xu1998invasion}%
  \BibitemOpen
  \bibfield  {author} {\bibinfo {author} {\bibfnamefont {B.}~\bibnamefont
  {Xu}}, \bibinfo {author} {\bibfnamefont {Y.}~\bibnamefont {Yortsos}},\ and\
  \bibinfo {author} {\bibfnamefont {D.}~\bibnamefont {Salin}},\ }\bibfield
  {title} {\bibinfo {title} {Invasion percolation with viscous forces},\
  }\href@noop {} {\bibfield  {journal} {\bibinfo  {journal} {Physical Review
  E}\ }\textbf {\bibinfo {volume} {57}},\ \bibinfo {pages} {739} (\bibinfo
  {year} {1998})}\BibitemShut {NoStop}%
\bibitem [{\citenamefont {Xu}\ \emph {et~al.}(2008)\citenamefont {Xu},
  \citenamefont {Davies}, \citenamefont {Schofield},\ and\ \citenamefont
  {Weitz}}]{xu2008dynamics}%
  \BibitemOpen
  \bibfield  {author} {\bibinfo {author} {\bibfnamefont {L.}~\bibnamefont
  {Xu}}, \bibinfo {author} {\bibfnamefont {S.}~\bibnamefont {Davies}}, \bibinfo
  {author} {\bibfnamefont {A.~B.}\ \bibnamefont {Schofield}},\ and\ \bibinfo
  {author} {\bibfnamefont {D.~A.}\ \bibnamefont {Weitz}},\ }\bibfield  {title}
  {\bibinfo {title} {Dynamics of drying in {3D} porous media},\ }\href@noop {}
  {\bibfield  {journal} {\bibinfo  {journal} {Physical Review Letters}\
  }\textbf {\bibinfo {volume} {101}},\ \bibinfo {pages} {094502} (\bibinfo
  {year} {2008})}\BibitemShut {NoStop}%
\bibitem [{\citenamefont {Joekar-Niasar}\ and\ \citenamefont
  {Hassanizadeh}(2012)}]{joekar2012analysis}%
  \BibitemOpen
  \bibfield  {author} {\bibinfo {author} {\bibfnamefont {V.}~\bibnamefont
  {Joekar-Niasar}}\ and\ \bibinfo {author} {\bibfnamefont {S.}~\bibnamefont
  {Hassanizadeh}},\ }\bibfield  {title} {\bibinfo {title} {Analysis of
  fundamentals of two-phase flow in porous media using dynamic pore-network
  models: A review},\ }\href@noop {} {\bibfield  {journal} {\bibinfo  {journal}
  {Critical Reviews in Environmental Science and Technology}\ }\textbf
  {\bibinfo {volume} {42}},\ \bibinfo {pages} {1895} (\bibinfo {year}
  {2012})}\BibitemShut {NoStop}%
\bibitem [{\citenamefont {Krummel}\ \emph {et~al.}(2013)\citenamefont
  {Krummel}, \citenamefont {Datta}, \citenamefont {M{\"u}nster},\ and\
  \citenamefont {Weitz}}]{krummel2013visualizing}%
  \BibitemOpen
  \bibfield  {author} {\bibinfo {author} {\bibfnamefont {A.~T.}\ \bibnamefont
  {Krummel}}, \bibinfo {author} {\bibfnamefont {S.~S.}\ \bibnamefont {Datta}},
  \bibinfo {author} {\bibfnamefont {S.}~\bibnamefont {M{\"u}nster}},\ and\
  \bibinfo {author} {\bibfnamefont {D.~A.}\ \bibnamefont {Weitz}},\ }\bibfield
  {title} {\bibinfo {title} {Visualizing multiphase flow and trapped fluid
  configurations in a model three-dimensional porous medium},\ }\href@noop {}
  {\bibfield  {journal} {\bibinfo  {journal} {AIChE Journal}\ }\textbf
  {\bibinfo {volume} {59}},\ \bibinfo {pages} {1022} (\bibinfo {year}
  {2013})}\BibitemShut {NoStop}%
\bibitem [{\citenamefont {Holtzman}\ and\ \citenamefont
  {Juanes}(2010)}]{holtzman2010crossover}%
  \BibitemOpen
  \bibfield  {author} {\bibinfo {author} {\bibfnamefont {R.}~\bibnamefont
  {Holtzman}}\ and\ \bibinfo {author} {\bibfnamefont {R.}~\bibnamefont
  {Juanes}},\ }\bibfield  {title} {\bibinfo {title} {Crossover from fingering
  to fracturing in deformable disordered media},\ }\href
  {https://doi.org/10.1103/PhysRevE.82.046305} {\bibfield  {journal} {\bibinfo
  {journal} {Physical Review E}\ }\textbf {\bibinfo {volume} {82}},\ \bibinfo
  {pages} {046305} (\bibinfo {year} {2010})}\BibitemShut {NoStop}%
\bibitem [{\citenamefont {Holtzman}\ \emph {et~al.}(2012)\citenamefont
  {Holtzman}, \citenamefont {Szulczewski},\ and\ \citenamefont
  {Juanes}}]{holtzman2012capillary}%
  \BibitemOpen
  \bibfield  {author} {\bibinfo {author} {\bibfnamefont {R.}~\bibnamefont
  {Holtzman}}, \bibinfo {author} {\bibfnamefont {M.~L.}\ \bibnamefont
  {Szulczewski}},\ and\ \bibinfo {author} {\bibfnamefont {R.}~\bibnamefont
  {Juanes}},\ }\bibfield  {title} {\bibinfo {title} {Capillary fracturing in
  granular media},\ }\href@noop {} {\bibfield  {journal} {\bibinfo  {journal}
  {Physical Review Letters}\ }\textbf {\bibinfo {volume} {108}},\ \bibinfo
  {pages} {264504} (\bibinfo {year} {2012})}\BibitemShut {NoStop}%
\bibitem [{\citenamefont {Derr}\ \emph {et~al.}(2020)\citenamefont {Derr},
  \citenamefont {Fronk}, \citenamefont {Weber}, \citenamefont {Mahadevan},
  \citenamefont {Rycroft},\ and\ \citenamefont {Mahadevan}}]{derr2020flow}%
  \BibitemOpen
  \bibfield  {author} {\bibinfo {author} {\bibfnamefont {N.~J.}\ \bibnamefont
  {Derr}}, \bibinfo {author} {\bibfnamefont {D.~C.}\ \bibnamefont {Fronk}},
  \bibinfo {author} {\bibfnamefont {C.~A.}\ \bibnamefont {Weber}}, \bibinfo
  {author} {\bibfnamefont {A.}~\bibnamefont {Mahadevan}}, \bibinfo {author}
  {\bibfnamefont {C.~H.}\ \bibnamefont {Rycroft}},\ and\ \bibinfo {author}
  {\bibfnamefont {L.}~\bibnamefont {Mahadevan}},\ }\bibfield  {title} {\bibinfo
  {title} {Flow-driven branching in a frangible porous medium},\ }\href@noop {}
  {\bibfield  {journal} {\bibinfo  {journal} {Physical Review Letters}\
  }\textbf {\bibinfo {volume} {125}},\ \bibinfo {pages} {158002} (\bibinfo
  {year} {2020})}\BibitemShut {NoStop}%
\bibitem [{\citenamefont {Style}\ \emph {et~al.}(2015)\citenamefont {Style},
  \citenamefont {Isa},\ and\ \citenamefont {Dufresne}}]{style2015adsorption}%
  \BibitemOpen
  \bibfield  {author} {\bibinfo {author} {\bibfnamefont {R.~W.}\ \bibnamefont
  {Style}}, \bibinfo {author} {\bibfnamefont {L.}~\bibnamefont {Isa}},\ and\
  \bibinfo {author} {\bibfnamefont {E.~R.}\ \bibnamefont {Dufresne}},\
  }\bibfield  {title} {\bibinfo {title} {Adsorption of soft particles at fluid
  interfaces},\ }\href@noop {} {\bibfield  {journal} {\bibinfo  {journal} {Soft
  Matter}\ }\textbf {\bibinfo {volume} {11}},\ \bibinfo {pages} {7412}
  (\bibinfo {year} {2015})}\BibitemShut {NoStop}%
\bibitem [{\citenamefont {Means}\ and\ \citenamefont
  {Wuayaratne}(1982)}]{means1982role}%
  \BibitemOpen
  \bibfield  {author} {\bibinfo {author} {\bibfnamefont {J.}~\bibnamefont
  {Means}}\ and\ \bibinfo {author} {\bibfnamefont {R.}~\bibnamefont
  {Wuayaratne}},\ }\bibfield  {title} {\bibinfo {title} {Role of natural
  colloids in the transport of hydrophobic pollutants},\ }\href@noop {}
  {\bibfield  {journal} {\bibinfo  {journal} {Science}\ }\textbf {\bibinfo
  {volume} {215}},\ \bibinfo {pages} {968} (\bibinfo {year}
  {1982})}\BibitemShut {NoStop}%
\bibitem [{\citenamefont {Tipping}\ and\ \citenamefont
  {Higgins}(1982)}]{tipping1982effect}%
  \BibitemOpen
  \bibfield  {author} {\bibinfo {author} {\bibfnamefont {E.}~\bibnamefont
  {Tipping}}\ and\ \bibinfo {author} {\bibfnamefont {D.}~\bibnamefont
  {Higgins}},\ }\bibfield  {title} {\bibinfo {title} {The effect of adsorbed
  humic substances on the colloid stability of haematite particles},\
  }\href@noop {} {\bibfield  {journal} {\bibinfo  {journal} {Colloids and
  Surfaces}\ }\textbf {\bibinfo {volume} {5}},\ \bibinfo {pages} {85} (\bibinfo
  {year} {1982})}\BibitemShut {NoStop}%
\bibitem [{\citenamefont {Gibbs}(1983)}]{gibbs1983effect}%
  \BibitemOpen
  \bibfield  {author} {\bibinfo {author} {\bibfnamefont {R.~J.}\ \bibnamefont
  {Gibbs}},\ }\bibfield  {title} {\bibinfo {title} {Effect of natural organic
  coatings on the coagulation of particles},\ }\href@noop {} {\bibfield
  {journal} {\bibinfo  {journal} {Environmental Science \& Technology}\
  }\textbf {\bibinfo {volume} {17}},\ \bibinfo {pages} {237} (\bibinfo {year}
  {1983})}\BibitemShut {NoStop}%
\bibitem [{\citenamefont {Corapcioglu}\ and\ \citenamefont
  {Jiang}(1993)}]{corapcioglu1993colloid}%
  \BibitemOpen
  \bibfield  {author} {\bibinfo {author} {\bibfnamefont {M.~Y.}\ \bibnamefont
  {Corapcioglu}}\ and\ \bibinfo {author} {\bibfnamefont {S.}~\bibnamefont
  {Jiang}},\ }\bibfield  {title} {\bibinfo {title} {Colloid-facilitated
  groundwater contaminant transport},\ }\href@noop {} {\bibfield  {journal}
  {\bibinfo  {journal} {Water Resources Research}\ }\textbf {\bibinfo {volume}
  {29}},\ \bibinfo {pages} {2215} (\bibinfo {year} {1993})}\BibitemShut
  {NoStop}%
\bibitem [{\citenamefont {Ouyang}\ \emph {et~al.}(1996)\citenamefont {Ouyang},
  \citenamefont {Shinde}, \citenamefont {Mansell},\ and\ \citenamefont
  {Harris}}]{ouyang1996colloid}%
  \BibitemOpen
  \bibfield  {author} {\bibinfo {author} {\bibfnamefont {Y.}~\bibnamefont
  {Ouyang}}, \bibinfo {author} {\bibfnamefont {D.}~\bibnamefont {Shinde}},
  \bibinfo {author} {\bibfnamefont {R.}~\bibnamefont {Mansell}},\ and\ \bibinfo
  {author} {\bibfnamefont {W.}~\bibnamefont {Harris}},\ }\bibfield  {title}
  {\bibinfo {title} {Colloid-enhanced transport of chemicals in subsurface
  environments: A review},\ }\href@noop {} {\bibfield  {journal} {\bibinfo
  {journal} {Critical Reviews in Environmental Science and Technology}\
  }\textbf {\bibinfo {volume} {26}},\ \bibinfo {pages} {189} (\bibinfo {year}
  {1996})}\BibitemShut {NoStop}%
\bibitem [{\citenamefont {Hendraningrat}\ \emph {et~al.}(2013)\citenamefont
  {Hendraningrat}, \citenamefont {Engeset}, \citenamefont {Suwarno},
  \citenamefont {Li},\ and\ \citenamefont
  {Tors{\ae}ter}}]{hendraningrat2013laboratory}%
  \BibitemOpen
  \bibfield  {author} {\bibinfo {author} {\bibfnamefont {L.}~\bibnamefont
  {Hendraningrat}}, \bibinfo {author} {\bibfnamefont {B.}~\bibnamefont
  {Engeset}}, \bibinfo {author} {\bibfnamefont {S.}~\bibnamefont {Suwarno}},
  \bibinfo {author} {\bibfnamefont {S.}~\bibnamefont {Li}},\ and\ \bibinfo
  {author} {\bibfnamefont {O.}~\bibnamefont {Tors{\ae}ter}},\ }\bibfield
  {title} {\bibinfo {title} {{Laboratory investigation of porosity and
  permeability impairment in Berea sandstones due to hydrophilic nanoparticle
  retention}},\ }in\ \href@noop {} {\emph {\bibinfo {booktitle} {Paper
  SCA2013-062 presented at the International Symposium of the Society of Core
  Analysts held in Napa Valley, California, USA}}}\ (\bibinfo {year} {2013})\
  pp.\ \bibinfo {pages} {16--19}\BibitemShut {NoStop}%
\bibitem [{\citenamefont {Feia}\ \emph {et~al.}(2015)\citenamefont {Feia},
  \citenamefont {Dupla}, \citenamefont {Ghabezloo}, \citenamefont {Sulem},
  \citenamefont {Canou}, \citenamefont {Onaisi}, \citenamefont {Lescanne},\
  and\ \citenamefont {Aubry}}]{feia2015experimental}%
  \BibitemOpen
  \bibfield  {author} {\bibinfo {author} {\bibfnamefont {S.}~\bibnamefont
  {Feia}}, \bibinfo {author} {\bibfnamefont {J.~C.}\ \bibnamefont {Dupla}},
  \bibinfo {author} {\bibfnamefont {S.}~\bibnamefont {Ghabezloo}}, \bibinfo
  {author} {\bibfnamefont {J.}~\bibnamefont {Sulem}}, \bibinfo {author}
  {\bibfnamefont {J.}~\bibnamefont {Canou}}, \bibinfo {author} {\bibfnamefont
  {A.}~\bibnamefont {Onaisi}}, \bibinfo {author} {\bibfnamefont
  {H.}~\bibnamefont {Lescanne}},\ and\ \bibinfo {author} {\bibfnamefont
  {E.}~\bibnamefont {Aubry}},\ }\bibfield  {title} {\bibinfo {title}
  {Experimental investigation of particle suspension injection and permeability
  impairment in porous media},\ }\href@noop {} {\bibfield  {journal} {\bibinfo
  {journal} {Geomechanics for Energy and the Environment}\ }\textbf {\bibinfo
  {volume} {3}},\ \bibinfo {pages} {24} (\bibinfo {year} {2015})}\BibitemShut
  {NoStop}%
\bibitem [{\citenamefont {Gerber}\ \emph {et~al.}(2019)\citenamefont {Gerber},
  \citenamefont {Bensouda}, \citenamefont {Weitz},\ and\ \citenamefont
  {Coussot}}]{gerber2019self}%
  \BibitemOpen
  \bibfield  {author} {\bibinfo {author} {\bibfnamefont {G.}~\bibnamefont
  {Gerber}}, \bibinfo {author} {\bibfnamefont {M.}~\bibnamefont {Bensouda}},
  \bibinfo {author} {\bibfnamefont {D.~A.}\ \bibnamefont {Weitz}},\ and\
  \bibinfo {author} {\bibfnamefont {P.}~\bibnamefont {Coussot}},\ }\bibfield
  {title} {\bibinfo {title} {Self-limited accumulation of colloids in porous
  media},\ }\href@noop {} {\bibfield  {journal} {\bibinfo  {journal} {Physical
  Review Letters}\ }\textbf {\bibinfo {volume} {123}},\ \bibinfo {pages}
  {158005} (\bibinfo {year} {2019})}\BibitemShut {NoStop}%
\bibitem [{\citenamefont {Bizmark}\ \emph {et~al.}(2020)\citenamefont
  {Bizmark}, \citenamefont {Schneider}, \citenamefont {Priestley},\ and\
  \citenamefont {Datta}}]{bizmark2020multiscale}%
  \BibitemOpen
  \bibfield  {author} {\bibinfo {author} {\bibfnamefont {N.}~\bibnamefont
  {Bizmark}}, \bibinfo {author} {\bibfnamefont {J.}~\bibnamefont {Schneider}},
  \bibinfo {author} {\bibfnamefont {R.~D.}\ \bibnamefont {Priestley}},\ and\
  \bibinfo {author} {\bibfnamefont {S.~S.}\ \bibnamefont {Datta}},\ }\bibfield
  {title} {\bibinfo {title} {Multiscale dynamics of colloidal deposition and
  erosion in porous media},\ }\href@noop {} {\bibfield  {journal} {\bibinfo
  {journal} {Science Advances}\ }\textbf {\bibinfo {volume} {6}},\ \bibinfo
  {pages} {eabc2530} (\bibinfo {year} {2020})}\BibitemShut {NoStop}%
\bibitem [{\citenamefont {Gerber}\ \emph {et~al.}(2020)\citenamefont {Gerber},
  \citenamefont {Weitz},\ and\ \citenamefont
  {Coussot}}]{gerber2020propagation}%
  \BibitemOpen
  \bibfield  {author} {\bibinfo {author} {\bibfnamefont {G.}~\bibnamefont
  {Gerber}}, \bibinfo {author} {\bibfnamefont {D.}~\bibnamefont {Weitz}},\ and\
  \bibinfo {author} {\bibfnamefont {P.}~\bibnamefont {Coussot}},\ }\bibfield
  {title} {\bibinfo {title} {{Propagation and adsorption of nanoparticles in
  porous medium as traveling waves}},\ }\href@noop {} {\bibfield  {journal}
  {\bibinfo  {journal} {Physical Review Research}\ }\textbf {\bibinfo {volume}
  {2}},\ \bibinfo {pages} {033074} (\bibinfo {year} {2020})}\BibitemShut
  {NoStop}%
\bibitem [{\citenamefont {Li}\ \emph {et~al.}(2020)\citenamefont {Li},
  \citenamefont {Su}, \citenamefont {Lv},\ and\ \citenamefont
  {Tu}}]{li2020asphaltene}%
  \BibitemOpen
  \bibfield  {author} {\bibinfo {author} {\bibfnamefont {L.}~\bibnamefont
  {Li}}, \bibinfo {author} {\bibfnamefont {Y.}~\bibnamefont {Su}}, \bibinfo
  {author} {\bibfnamefont {Y.}~\bibnamefont {Lv}},\ and\ \bibinfo {author}
  {\bibfnamefont {J.}~\bibnamefont {Tu}},\ }\bibfield  {title} {\bibinfo
  {title} {Asphaltene deposition and permeability impairment in shale
  reservoirs during {CO$_2$} huff-n-puff {EOR} process},\ }\href@noop {}
  {\bibfield  {journal} {\bibinfo  {journal} {Petroleum Science and
  Technology}\ }\textbf {\bibinfo {volume} {38}},\ \bibinfo {pages} {384}
  (\bibinfo {year} {2020})}\BibitemShut {NoStop}%
\bibitem [{\citenamefont {McCarthy}\ and\ \citenamefont
  {Zachara}(1989)}]{mccarthy1989t}%
  \BibitemOpen
  \bibfield  {author} {\bibinfo {author} {\bibfnamefont {J.}~\bibnamefont
  {McCarthy}}\ and\ \bibinfo {author} {\bibfnamefont {J.}~\bibnamefont
  {Zachara}},\ }\bibfield  {title} {\bibinfo {title} {{ES\&T} features:
  Subsurface transport of contaminants},\ }\href@noop {} {\bibfield  {journal}
  {\bibinfo  {journal} {Environmental Science \& Technology}\ }\textbf
  {\bibinfo {volume} {23}},\ \bibinfo {pages} {496} (\bibinfo {year}
  {1989})}\BibitemShut {NoStop}%
\bibitem [{\citenamefont {Kan}\ and\ \citenamefont
  {Tomson}(1990)}]{kan1990ground}%
  \BibitemOpen
  \bibfield  {author} {\bibinfo {author} {\bibfnamefont {A.~T.}\ \bibnamefont
  {Kan}}\ and\ \bibinfo {author} {\bibfnamefont {M.~B.}\ \bibnamefont
  {Tomson}},\ }\bibfield  {title} {\bibinfo {title} {Ground water transport of
  hydrophobic organic compounds in the presence of dissolved organic matter},\
  }\href@noop {} {\bibfield  {journal} {\bibinfo  {journal} {Environmental
  Toxicology and Chemistry: An International Journal}\ }\textbf {\bibinfo
  {volume} {9}},\ \bibinfo {pages} {253} (\bibinfo {year} {1990})}\BibitemShut
  {NoStop}%
\bibitem [{\citenamefont {Johnson}\ and\ \citenamefont
  {Logan}(1996)}]{johnson1996enhanced}%
  \BibitemOpen
  \bibfield  {author} {\bibinfo {author} {\bibfnamefont {W.~P.}\ \bibnamefont
  {Johnson}}\ and\ \bibinfo {author} {\bibfnamefont {B.~E.}\ \bibnamefont
  {Logan}},\ }\bibfield  {title} {\bibinfo {title} {Enhanced transport of
  bacteria in porous media by sediment-phase and aqueous-phase natural organic
  matter},\ }\href@noop {} {\bibfield  {journal} {\bibinfo  {journal} {Water
  Research}\ }\textbf {\bibinfo {volume} {30}},\ \bibinfo {pages} {923}
  (\bibinfo {year} {1996})}\BibitemShut {NoStop}%
\bibitem [{\citenamefont {Franchi}\ and\ \citenamefont
  {O'Melia}(2003)}]{franchi2003effects}%
  \BibitemOpen
  \bibfield  {author} {\bibinfo {author} {\bibfnamefont {A.}~\bibnamefont
  {Franchi}}\ and\ \bibinfo {author} {\bibfnamefont {C.~R.}\ \bibnamefont
  {O'Melia}},\ }\bibfield  {title} {\bibinfo {title} {Effects of natural
  organic matter and solution chemistry on the deposition and reentrainment of
  colloids in porous media},\ }\href@noop {} {\bibfield  {journal} {\bibinfo
  {journal} {Environmental Science \& Technology}\ }\textbf {\bibinfo {volume}
  {37}},\ \bibinfo {pages} {1122} (\bibinfo {year} {2003})}\BibitemShut
  {NoStop}%
\bibitem [{\citenamefont {Pelley}\ and\ \citenamefont
  {Tufenkji}(2008)}]{pelley2008effect}%
  \BibitemOpen
  \bibfield  {author} {\bibinfo {author} {\bibfnamefont {A.~J.}\ \bibnamefont
  {Pelley}}\ and\ \bibinfo {author} {\bibfnamefont {N.}~\bibnamefont
  {Tufenkji}},\ }\bibfield  {title} {\bibinfo {title} {Effect of particle size
  and natural organic matter on the migration of nano-and microscale latex
  particles in saturated porous media},\ }\href@noop {} {\bibfield  {journal}
  {\bibinfo  {journal} {Journal of Colloid and Interface Science}\ }\textbf
  {\bibinfo {volume} {321}},\ \bibinfo {pages} {74} (\bibinfo {year}
  {2008})}\BibitemShut {NoStop}%
\bibitem [{\citenamefont {Wilkinson}(1986)}]{wilkinson1986percolation}%
  \BibitemOpen
  \bibfield  {author} {\bibinfo {author} {\bibfnamefont {D.}~\bibnamefont
  {Wilkinson}},\ }\bibfield  {title} {\bibinfo {title} {Percolation effects in
  immiscible displacement},\ }\href@noop {} {\bibfield  {journal} {\bibinfo
  {journal} {Physical Review A}\ }\textbf {\bibinfo {volume} {34}},\ \bibinfo
  {pages} {1380} (\bibinfo {year} {1986})}\BibitemShut {NoStop}%
\bibitem [{\citenamefont {Birovljev}\ \emph {et~al.}(1991)\citenamefont
  {Birovljev}, \citenamefont {Furuberg}, \citenamefont {Feder}, \citenamefont
  {Jssang}, \citenamefont {Mly},\ and\ \citenamefont
  {Aharony}}]{birovljev1991gravity}%
  \BibitemOpen
  \bibfield  {author} {\bibinfo {author} {\bibfnamefont {A.}~\bibnamefont
  {Birovljev}}, \bibinfo {author} {\bibfnamefont {L.}~\bibnamefont {Furuberg}},
  \bibinfo {author} {\bibfnamefont {J.}~\bibnamefont {Feder}}, \bibinfo
  {author} {\bibfnamefont {T.}~\bibnamefont {Jssang}}, \bibinfo {author}
  {\bibfnamefont {K.}~\bibnamefont {Mly}},\ and\ \bibinfo {author}
  {\bibfnamefont {A.}~\bibnamefont {Aharony}},\ }\bibfield  {title} {\bibinfo
  {title} {Gravity invasion percolation in two dimensions: Experiment and
  simulation},\ }\href@noop {} {\bibfield  {journal} {\bibinfo  {journal}
  {Physical Review Letters}\ }\textbf {\bibinfo {volume} {67}},\ \bibinfo
  {pages} {584} (\bibinfo {year} {1991})}\BibitemShut {NoStop}%
\bibitem [{\citenamefont {Masson}(2016)}]{masson2016fast}%
  \BibitemOpen
  \bibfield  {author} {\bibinfo {author} {\bibfnamefont {Y.}~\bibnamefont
  {Masson}},\ }\bibfield  {title} {\bibinfo {title} {A fast two-step algorithm
  for invasion percolation with trapping},\ }\href@noop {} {\bibfield
  {journal} {\bibinfo  {journal} {Computers \& Geosciences}\ }\textbf {\bibinfo
  {volume} {90}},\ \bibinfo {pages} {41} (\bibinfo {year} {2016})}\BibitemShut
  {NoStop}%
\bibitem [{\citenamefont {J{\"a}ger}\ \emph {et~al.}(2017)\citenamefont
  {J{\"a}ger}, \citenamefont {Mendoza},\ and\ \citenamefont
  {Herrmann}}]{jager2017channelization}%
  \BibitemOpen
  \bibfield  {author} {\bibinfo {author} {\bibfnamefont {R.}~\bibnamefont
  {J{\"a}ger}}, \bibinfo {author} {\bibfnamefont {M.}~\bibnamefont {Mendoza}},\
  and\ \bibinfo {author} {\bibfnamefont {H.~J.}\ \bibnamefont {Herrmann}},\
  }\bibfield  {title} {\bibinfo {title} {Channelization in porous media driven
  by erosion and deposition},\ }\href@noop {} {\bibfield  {journal} {\bibinfo
  {journal} {Physical Review E}\ }\textbf {\bibinfo {volume} {95}},\ \bibinfo
  {pages} {013110} (\bibinfo {year} {2017})}\BibitemShut {NoStop}%
\bibitem [{\citenamefont {Khodaparast}\ \emph {et~al.}(2017)\citenamefont
  {Khodaparast}, \citenamefont {Kim}, \citenamefont {Silpe},\ and\
  \citenamefont {Stone}}]{khodaparast2017bubble}%
  \BibitemOpen
  \bibfield  {author} {\bibinfo {author} {\bibfnamefont {S.}~\bibnamefont
  {Khodaparast}}, \bibinfo {author} {\bibfnamefont {M.~K.}\ \bibnamefont
  {Kim}}, \bibinfo {author} {\bibfnamefont {J.~E.}\ \bibnamefont {Silpe}},\
  and\ \bibinfo {author} {\bibfnamefont {H.~A.}\ \bibnamefont {Stone}},\
  }\bibfield  {title} {\bibinfo {title} {Bubble-driven detachment of bacteria
  from confined microgeometries},\ }\href@noop {} {\bibfield  {journal}
  {\bibinfo  {journal} {Environmental Science \& Technology}\ }\textbf
  {\bibinfo {volume} {51}},\ \bibinfo {pages} {1340} (\bibinfo {year}
  {2017})}\BibitemShut {NoStop}%
\bibitem [{\citenamefont {Yu}\ \emph {et~al.}(2017)\citenamefont {Yu},
  \citenamefont {Khodaparast},\ and\ \citenamefont {Stone}}]{yu2017armoring}%
  \BibitemOpen
  \bibfield  {author} {\bibinfo {author} {\bibfnamefont {Y.~E.}\ \bibnamefont
  {Yu}}, \bibinfo {author} {\bibfnamefont {S.}~\bibnamefont {Khodaparast}},\
  and\ \bibinfo {author} {\bibfnamefont {H.~A.}\ \bibnamefont {Stone}},\
  }\bibfield  {title} {\bibinfo {title} {Armoring confined bubbles in the flow
  of colloidal suspensions},\ }\href@noop {} {\bibfield  {journal} {\bibinfo
  {journal} {Soft Matter}\ }\textbf {\bibinfo {volume} {13}},\ \bibinfo {pages}
  {2857} (\bibinfo {year} {2017})}\BibitemShut {NoStop}%
\bibitem [{\citenamefont {Yin}\ \emph {et~al.}(2018)\citenamefont {Yin},
  \citenamefont {Shin}, \citenamefont {Frechette}, \citenamefont {Colosqui},\
  and\ \citenamefont {Drazer}}]{yin2018dynamic}%
  \BibitemOpen
  \bibfield  {author} {\bibinfo {author} {\bibfnamefont {T.}~\bibnamefont
  {Yin}}, \bibinfo {author} {\bibfnamefont {D.}~\bibnamefont {Shin}}, \bibinfo
  {author} {\bibfnamefont {J.}~\bibnamefont {Frechette}}, \bibinfo {author}
  {\bibfnamefont {C.~E.}\ \bibnamefont {Colosqui}},\ and\ \bibinfo {author}
  {\bibfnamefont {G.}~\bibnamefont {Drazer}},\ }\bibfield  {title} {\bibinfo
  {title} {Dynamic effects on the mobilization of a deposited nanoparticle by a
  moving liquid-liquid interface},\ }\href@noop {} {\bibfield  {journal}
  {\bibinfo  {journal} {Physical Review Letters}\ }\textbf {\bibinfo {volume}
  {121}},\ \bibinfo {pages} {238002} (\bibinfo {year} {2018})}\BibitemShut
  {NoStop}%
\bibitem [{\citenamefont {Jeong}\ \emph {et~al.}(2022)\citenamefont {Jeong},
  \citenamefont {Xing}, \citenamefont {Boutin},\ and\ \citenamefont
  {Sauret}}]{jeong2022particulate}%
  \BibitemOpen
  \bibfield  {author} {\bibinfo {author} {\bibfnamefont {D.-H.}\ \bibnamefont
  {Jeong}}, \bibinfo {author} {\bibfnamefont {L.}~\bibnamefont {Xing}},
  \bibinfo {author} {\bibfnamefont {J.-B.}\ \bibnamefont {Boutin}},\ and\
  \bibinfo {author} {\bibfnamefont {A.}~\bibnamefont {Sauret}},\ }\bibfield
  {title} {\bibinfo {title} {Particulate suspension coating of capillary
  tubes},\ }\href@noop {} {\bibfield  {journal} {\bibinfo  {journal} {Soft
  Matter}\ }\textbf {\bibinfo {volume} {18}},\ \bibinfo {pages} {8124}
  (\bibinfo {year} {2022})}\BibitemShut {NoStop}%
\bibitem [{SI()}]{SI}%
  \BibitemOpen
  \href@noop {} {}\bibinfo {note} {See Supplemental Material at [URL will be
  inserted by publisher] for supporting calculations, figures, and
  movies.}\BibitemShut {Stop}%
\bibitem [{\citenamefont {Niemeyer}\ \emph {et~al.}(1984)\citenamefont
  {Niemeyer}, \citenamefont {Pietronero},\ and\ \citenamefont
  {Wiesmann}}]{niemeyer1984fractal}%
  \BibitemOpen
  \bibfield  {author} {\bibinfo {author} {\bibfnamefont {L.}~\bibnamefont
  {Niemeyer}}, \bibinfo {author} {\bibfnamefont {L.}~\bibnamefont
  {Pietronero}},\ and\ \bibinfo {author} {\bibfnamefont {H.~J.}\ \bibnamefont
  {Wiesmann}},\ }\bibfield  {title} {\bibinfo {title} {Fractal dimension of
  dielectric breakdown},\ }\href {https://doi.org/10.1103/PhysRevLett.52.1033}
  {\bibfield  {journal} {\bibinfo  {journal} {Physical Review Letters}\
  }\textbf {\bibinfo {volume} {52}},\ \bibinfo {pages} {1033} (\bibinfo {year}
  {1984})}\BibitemShut {NoStop}%
\bibitem [{\citenamefont {Lenormand}\ \emph {et~al.}(1988)\citenamefont
  {Lenormand}, \citenamefont {Touboul},\ and\ \citenamefont
  {Zarcone}}]{lenormand1988numerical}%
  \BibitemOpen
  \bibfield  {author} {\bibinfo {author} {\bibfnamefont {R.}~\bibnamefont
  {Lenormand}}, \bibinfo {author} {\bibfnamefont {E.}~\bibnamefont {Touboul}},\
  and\ \bibinfo {author} {\bibfnamefont {C.}~\bibnamefont {Zarcone}},\
  }\bibfield  {title} {\bibinfo {title} {Numerical models and experiments on
  immiscible displacements in porous media},\ }\href@noop {} {\bibfield
  {journal} {\bibinfo  {journal} {Journal of Fluid Mechanics}\ }\textbf
  {\bibinfo {volume} {189}},\ \bibinfo {pages} {165} (\bibinfo {year}
  {1988})}\BibitemShut {NoStop}%
\bibitem [{\citenamefont {Meakin}\ \emph {et~al.}(1992)\citenamefont {Meakin},
  \citenamefont {Feder}, \citenamefont {Frette},\ and\ \citenamefont
  {J{\o}ssang}}]{meakin1992invasion}%
  \BibitemOpen
  \bibfield  {author} {\bibinfo {author} {\bibfnamefont {P.}~\bibnamefont
  {Meakin}}, \bibinfo {author} {\bibfnamefont {J.}~\bibnamefont {Feder}},
  \bibinfo {author} {\bibfnamefont {V.}~\bibnamefont {Frette}},\ and\ \bibinfo
  {author} {\bibfnamefont {T.}~\bibnamefont {J{\o}ssang}},\ }\bibfield  {title}
  {\bibinfo {title} {Invasion percolation in a destabilizing gradient},\
  }\href@noop {} {\bibfield  {journal} {\bibinfo  {journal} {Physical Review
  A}\ }\textbf {\bibinfo {volume} {46}},\ \bibinfo {pages} {3357} (\bibinfo
  {year} {1992})}\BibitemShut {NoStop}%
\bibitem [{\citenamefont {Onody}\ \emph {et~al.}(1995)\citenamefont {Onody},
  \citenamefont {Posadas},\ and\ \citenamefont
  {Crestana}}]{onody1995experimental}%
  \BibitemOpen
  \bibfield  {author} {\bibinfo {author} {\bibfnamefont {R.~N.}\ \bibnamefont
  {Onody}}, \bibinfo {author} {\bibfnamefont {A.}~\bibnamefont {Posadas}},\
  and\ \bibinfo {author} {\bibfnamefont {S.}~\bibnamefont {Crestana}},\
  }\bibfield  {title} {\bibinfo {title} {Experimental studies of the fingering
  phenomena in two dimensions and simulation using a modified invasion
  percolation model},\ }\href@noop {} {\bibfield  {journal} {\bibinfo
  {journal} {Journal of Applied Physics}\ }\textbf {\bibinfo {volume} {78}},\
  \bibinfo {pages} {2970} (\bibinfo {year} {1995})}\BibitemShut {NoStop}%
\bibitem [{\citenamefont {Al-Housseiny}\ \emph {et~al.}(2012)\citenamefont
  {Al-Housseiny}, \citenamefont {Tsai},\ and\ \citenamefont
  {Stone}}]{al2012control}%
  \BibitemOpen
  \bibfield  {author} {\bibinfo {author} {\bibfnamefont {T.~T.}\ \bibnamefont
  {Al-Housseiny}}, \bibinfo {author} {\bibfnamefont {P.~A.}\ \bibnamefont
  {Tsai}},\ and\ \bibinfo {author} {\bibfnamefont {H.~A.}\ \bibnamefont
  {Stone}},\ }\bibfield  {title} {\bibinfo {title} {Control of interfacial
  instabilities using flow geometry},\ }\href@noop {} {\bibfield  {journal}
  {\bibinfo  {journal} {Nature Physics}\ }\textbf {\bibinfo {volume} {8}},\
  \bibinfo {pages} {747} (\bibinfo {year} {2012})}\BibitemShut {NoStop}%
\bibitem [{\citenamefont {Datta}\ and\ \citenamefont
  {Weitz}(2013)}]{datta2013drainage}%
  \BibitemOpen
  \bibfield  {author} {\bibinfo {author} {\bibfnamefont {S.~S.}\ \bibnamefont
  {Datta}}\ and\ \bibinfo {author} {\bibfnamefont {D.~A.}\ \bibnamefont
  {Weitz}},\ }\bibfield  {title} {\bibinfo {title} {Drainage in a model
  stratified porous medium},\ }\href@noop {} {\bibfield  {journal} {\bibinfo
  {journal} {Europhysics Letters}\ }\textbf {\bibinfo {volume} {101}},\
  \bibinfo {pages} {14002} (\bibinfo {year} {2013})}\BibitemShut {NoStop}%
\bibitem [{\citenamefont {Jackson}\ \emph {et~al.}(2017)\citenamefont
  {Jackson}, \citenamefont {Power}, \citenamefont {Giddings},\ and\
  \citenamefont {Stevens}}]{jackson2017stability}%
  \BibitemOpen
  \bibfield  {author} {\bibinfo {author} {\bibfnamefont {S.}~\bibnamefont
  {Jackson}}, \bibinfo {author} {\bibfnamefont {H.}~\bibnamefont {Power}},
  \bibinfo {author} {\bibfnamefont {D.}~\bibnamefont {Giddings}},\ and\
  \bibinfo {author} {\bibfnamefont {D.}~\bibnamefont {Stevens}},\ }\bibfield
  {title} {\bibinfo {title} {The stability of immiscible viscous fingering in
  hele-shaw cells with spatially varying permeability},\ }\href@noop {}
  {\bibfield  {journal} {\bibinfo  {journal} {Computer Methods in Applied
  Mechanics and Engineering}\ }\textbf {\bibinfo {volume} {320}},\ \bibinfo
  {pages} {606} (\bibinfo {year} {2017})}\BibitemShut {NoStop}%
\bibitem [{\citenamefont {Biswas}\ \emph {et~al.}(2018)\citenamefont {Biswas},
  \citenamefont {Fantinel}, \citenamefont {Borgman}, \citenamefont {Holtzman},\
  and\ \citenamefont {Goehring}}]{biswas2018drying}%
  \BibitemOpen
  \bibfield  {author} {\bibinfo {author} {\bibfnamefont {S.}~\bibnamefont
  {Biswas}}, \bibinfo {author} {\bibfnamefont {P.}~\bibnamefont {Fantinel}},
  \bibinfo {author} {\bibfnamefont {O.}~\bibnamefont {Borgman}}, \bibinfo
  {author} {\bibfnamefont {R.}~\bibnamefont {Holtzman}},\ and\ \bibinfo
  {author} {\bibfnamefont {L.}~\bibnamefont {Goehring}},\ }\bibfield  {title}
  {\bibinfo {title} {Drying and percolation in correlated porous media},\
  }\href {https://doi.org/10.1103/PhysRevFluids.3.124307} {\bibfield  {journal}
  {\bibinfo  {journal} {Physical Review Fluids}\ }\textbf {\bibinfo {volume}
  {3}},\ \bibinfo {pages} {124307} (\bibinfo {year} {2018})}\BibitemShut
  {NoStop}%
\bibitem [{\citenamefont {Lu}\ \emph {et~al.}(2019)\citenamefont {Lu},
  \citenamefont {Browne}, \citenamefont {Amchin}, \citenamefont {Nunes},\ and\
  \citenamefont {Datta}}]{lu2019controlling}%
  \BibitemOpen
  \bibfield  {author} {\bibinfo {author} {\bibfnamefont {N.~B.}\ \bibnamefont
  {Lu}}, \bibinfo {author} {\bibfnamefont {C.~A.}\ \bibnamefont {Browne}},
  \bibinfo {author} {\bibfnamefont {D.~B.}\ \bibnamefont {Amchin}}, \bibinfo
  {author} {\bibfnamefont {J.~K.}\ \bibnamefont {Nunes}},\ and\ \bibinfo
  {author} {\bibfnamefont {S.~S.}\ \bibnamefont {Datta}},\ }\bibfield  {title}
  {\bibinfo {title} {Controlling capillary fingering using pore size gradients
  in disordered media},\ }\href@noop {} {\bibfield  {journal} {\bibinfo
  {journal} {Physical Review Fluids}\ }\textbf {\bibinfo {volume} {4}},\
  \bibinfo {pages} {084303} (\bibinfo {year} {2019})}\BibitemShut {NoStop}%
\bibitem [{\citenamefont {Lu}\ \emph {et~al.}(2020)\citenamefont {Lu},
  \citenamefont {Pahlavan}, \citenamefont {Browne}, \citenamefont {Amchin},
  \citenamefont {Stone},\ and\ \citenamefont {Datta}}]{lu2020forced}%
  \BibitemOpen
  \bibfield  {author} {\bibinfo {author} {\bibfnamefont {N.~B.}\ \bibnamefont
  {Lu}}, \bibinfo {author} {\bibfnamefont {A.~A.}\ \bibnamefont {Pahlavan}},
  \bibinfo {author} {\bibfnamefont {C.~A.}\ \bibnamefont {Browne}}, \bibinfo
  {author} {\bibfnamefont {D.~B.}\ \bibnamefont {Amchin}}, \bibinfo {author}
  {\bibfnamefont {H.~A.}\ \bibnamefont {Stone}},\ and\ \bibinfo {author}
  {\bibfnamefont {S.~S.}\ \bibnamefont {Datta}},\ }\bibfield  {title} {\bibinfo
  {title} {Forced imbibition in stratified porous media},\ }\href@noop {}
  {\bibfield  {journal} {\bibinfo  {journal} {Physical Review Applied}\
  }\textbf {\bibinfo {volume} {14}},\ \bibinfo {pages} {054009} (\bibinfo
  {year} {2020})}\BibitemShut {NoStop}%
\bibitem [{\citenamefont {Lu}\ \emph {et~al.}(2021)\citenamefont {Lu},
  \citenamefont {Amchin},\ and\ \citenamefont {Datta}}]{lu2021forced}%
  \BibitemOpen
  \bibfield  {author} {\bibinfo {author} {\bibfnamefont {N.~B.}\ \bibnamefont
  {Lu}}, \bibinfo {author} {\bibfnamefont {D.~B.}\ \bibnamefont {Amchin}},\
  and\ \bibinfo {author} {\bibfnamefont {S.~S.}\ \bibnamefont {Datta}},\
  }\bibfield  {title} {\bibinfo {title} {Forced imbibition in stratified porous
  media: Fluid dynamics and breakthrough saturation},\ }\href@noop {}
  {\bibfield  {journal} {\bibinfo  {journal} {Physical Review Fluids}\ }\textbf
  {\bibinfo {volume} {6}},\ \bibinfo {pages} {114007} (\bibinfo {year}
  {2021})}\BibitemShut {NoStop}%
\bibitem [{\citenamefont {Jerauld}\ and\ \citenamefont
  {Salter}(1990)}]{jerauld1990effect}%
  \BibitemOpen
  \bibfield  {author} {\bibinfo {author} {\bibfnamefont {G.}~\bibnamefont
  {Jerauld}}\ and\ \bibinfo {author} {\bibfnamefont {S.}~\bibnamefont
  {Salter}},\ }\bibfield  {title} {\bibinfo {title} {The effect of
  pore-structure on hysteresis in relative permeability and capillary pressure:
  pore-level modeling},\ }\href@noop {} {\bibfield  {journal} {\bibinfo
  {journal} {Transport in Porous Media}\ }\textbf {\bibinfo {volume} {5}},\
  \bibinfo {pages} {103} (\bibinfo {year} {1990})}\BibitemShut {NoStop}%
\end{thebibliography}
\end{document}